\documentclass[conference]{IEEEtran}
\IEEEoverridecommandlockouts
\usepackage{amsmath,graphicx}
\usepackage{subcaption}
\usepackage{amsfonts} 

\usepackage{multirow,epstopdf}
\usepackage{booktabs}
\usepackage{algorithm2e}
\usepackage{amsmath}
\usepackage{bbm}
\usepackage{enumerate}
\usepackage[table,xcdraw]{xcolor}

\renewcommand{\baselinestretch}{0.965}

\def\BibTeX{{\rm B\kern-.05em{\sc i\kern-.025em b}\kern-.08em
    T\kern-.1667em\lower.7ex\hbox{E}\kern-.125emX}}
\begin{document}

\title{Ed-Fed: A generic federated learning framework with resource-aware client selection for edge devices\\
}

\author{\IEEEauthorblockN{Zitha Sasindran, Harsha Yelchuri, T. V. Prabhakar}
\IEEEauthorblockA{\textit{Department of Electronic Systems Engineering}\\
\textit{Indian Institute of Science}\\
Bengaluru, India, 560012.\\
Email: \{zithas, harshay, tvprabs\}@iisc.ac.in}}

\maketitle

\begin{abstract}
Federated learning (FL) has evolved as a prominent method for edge devices to cooperatively create a unified prediction model while securing their sensitive training data local to the device. Despite the existence of numerous research frameworks for simulating FL algorithms, they do not facilitate comprehensive deployment for automatic speech recognition tasks on heterogeneous edge devices. This is where Ed-Fed, a comprehensive and generic FL framework, comes in as a foundation for future practical FL system research. We also propose a novel resource-aware client selection algorithm to optimise the waiting time in the FL settings. We show that our approach can handle the straggler devices and dynamically set the training time for the selected devices in a round. Our evaluation has shown that the proposed approach significantly optimises waiting time in FL compared to conventional random client selection methods.
\end{abstract}

% \begin{IEEEkeywords}
% component, formatting, style, styling, insert
% \end{IEEEkeywords}

\section{Introduction}
With the advances in hardware and software technologies, edge devices are becoming increasingly powerful and intelligent. This enables the researchers to bring machine intelligence from cloud-based data centres to edge devices such as mobile phones. For instance, Google Pixel's ``Now Playing" option which allows us to recognise any song without internet and ``recorder" app with on-device speaker diarization show how powerful the mobile phones are becoming. Federated learning (FL) \cite{FL} introduces an additional dimension in the machine learning community in which models are collaboratively trained on edge devices, with the data remaining local to the device, in contrast to the centralised training approach, and hence ensures the data privacy of the users.
A general FL setting consists of a set of clients (edge devices) and a server. The server randomly chooses a subset of the devices from the available ones that want to take part in the FL round. A copy of the global model, which the server maintains, is first sent to this subset of clients. These clients use their data in the device to train models locally, and then send the updated weights back to the server. In conventional FL, the server cannot move on to the next step until it has received updates from every client. As a result, FL performance is constrained by the variability in waiting time experienced by each client due to model training on the device or communication delay during the transfer of model weights to the server.
These slowdowns caused by the clients with weaker network connections or limited computation capabilities is known as ``straggler effect". Once the server gets the updates from all the selected clients, the weights are aggregated using a strategy. This process is repeated for several rounds until the global model achieves the desired  accuracy. 

The random client selection approach for FL works well when there are no straggler devices. Prioritising for resource rich devices every time will result in the inability of straggler devices to participate  in the FL process, and can lead to a loss of generalisation in the global model and fairness in the learning process. To address this challenge, a more sophisticated client selection algorithm is needed that considers the presence of stragglers while minimising waiting time. Hence the algorithm should aim to balance the participation of straggler devices with less waiting time and fairness in the FL process. Current FL frameworks face several challenges when it comes to deployment on edge devices such as mobile phones. These devices often have limited memory and computational resources, which makes it difficult to store and execute complex models.  To overcome the existing challenges and make FL a practical and scalable solution for edge devices, it is important to develop FL frameworks that are designed specifically for these devices, taking into account their computational, privacy, and power constraints.

We provide a methodology for training the models on the device, along with an efficient client selection algorithm to handle straggler devices and FL functionalities, allowing them to be deployed on edge devices for FL settings. We also monitor the client devices' resources to ensure that they continue to function seamlessly even in dynamic environments.
Since this work is related to client selection, we do not consider asynchronous federated learning approaches. 
In this paper, we focus on the use case of automatic speech recognition (ASR), demonstrating how our FL framework can be used to improve the accuracy and robustness of speech recognition models.
We summarise our contributions as follows:
\begin{itemize}
    \item We present Ed-Fed, an end-to-end FL framework for edge devices with a resource-aware time-optimised client selection algorithm.
    \item We provide a complete methodology for training entire or fine tuned models on mobile phones with support for FL related functionalities.
    \item We formulate a client selection algorithm by considering the computation, storage, power, and phone-specific capabilities of the client devices with ability to handle stragglers, optimise the waiting time and adaptively assign the training time for devices based on the these information.
    \item We demonstrate the implementation with deployments on multiple mobile phones  to quantify the waiting time in client selection. 
    \item We present an extensive evaluation of our framework on both simulations and mobile phone settings using a custom created audio corpus with a heterogeneous set of speakers. 
\end{itemize}

This paper is organised as follows. We provide a brief literature survey in Section II. Then, we discuss briefly about our Ed-Fed framework for FL settings in Section III. In Section IV, we discuss our proposed resource-aware client selection algorithm used in the server, followed by the framework evaluation in Section V and results in Section VI. Finally, concluding remarks are presented in Section VII.

\section{Related works}
\textbf{FL Framework }Recently, there has been a lot of interest in training ASR models in FL settings \cite{flasr_wer,flasr_ibm, flasr_3, flasr_4,flasr_5}. Implementing ASR for federated settings in mobile phones is a challenge in itself due to the complex model architectures and dynamic nature of the speech signals, as well as the limited resources available in mobile devices.
Furthermore, due to factors such as noisy backgrounds, multiple accented speech, and different voice characteristics such as gender, pitch, phonation, loudness and tempo, the local speech data available on devices are non-IID in nature. This further complicates the training of ASR models. Several FL frameworks, including TensorFlow-Federated (TFF) \cite{tff}, and LEAF \cite{leaf}, support only the simulation of FL systems and do not propose an edge device deployment. Flower framework \cite{flower}, on the other hand, supports extending FL settings to edge devices. However, we can see that the ASR task with the flower framework is also limited to system-level simulations \cite{flasr_wer}. Moreover, they provide only transfer learning techniques \cite{flower_old} but not entire retraining of the model from scratch.  To the best of our knowledge, no existing framework has provided a comprehensive implementation of ASR on-device training, as well as FL settings for mobile phones.

\textbf{Client selection }  We focus mainly on bandit based client selection techniques in FL. In \cite{oort}, the authors formulated the client selection as a traditional multi-arm bandit based problem to select clients with better quality of data to improve the FL accuracy. The authors proposed upper confidence bound based client selection in \cite{csucb,mabcs} with the goal of reducing the overall time consumption of FL training including transmission time and local computation time. In \cite{birdsoffeathers}, the authors intelligently select clients by exploiting the data correlations among clients to improve FL learning performance. However, none of these studies used actual computation resource information from the devices to explain training latency. Meanwhile, \cite{optimisedadaptiveFL} took a similar approach to ours, taking into account resource information and grouping clients accordingly.  Even though the clients with similar resources are selected together, the need for adaptive setting training epochs is needed to control the stragglers. Furthermore, they did not take the battery information into account, which is critical.
\vspace{-0.1cm}

\section{Ed-Fed framework}
\label{sec:Ed-Fed framework}
This section provides a detailed overview of our Ed-Fed framework.
% The key components present in the Ed-Fed framework is depicted are Figure \ref{fig:flframework}.
We first discuss the methodology for facilitating on-device training and weight updation of models in the clients, followed by a brief overview on the communication protocol and the server-side algorithms  used.

\subsection{Client}
\label{sec:clients}
\begin{figure}[!t]
    \centering
	\includegraphics[width = 0.45\textwidth,height = 4.5cm]{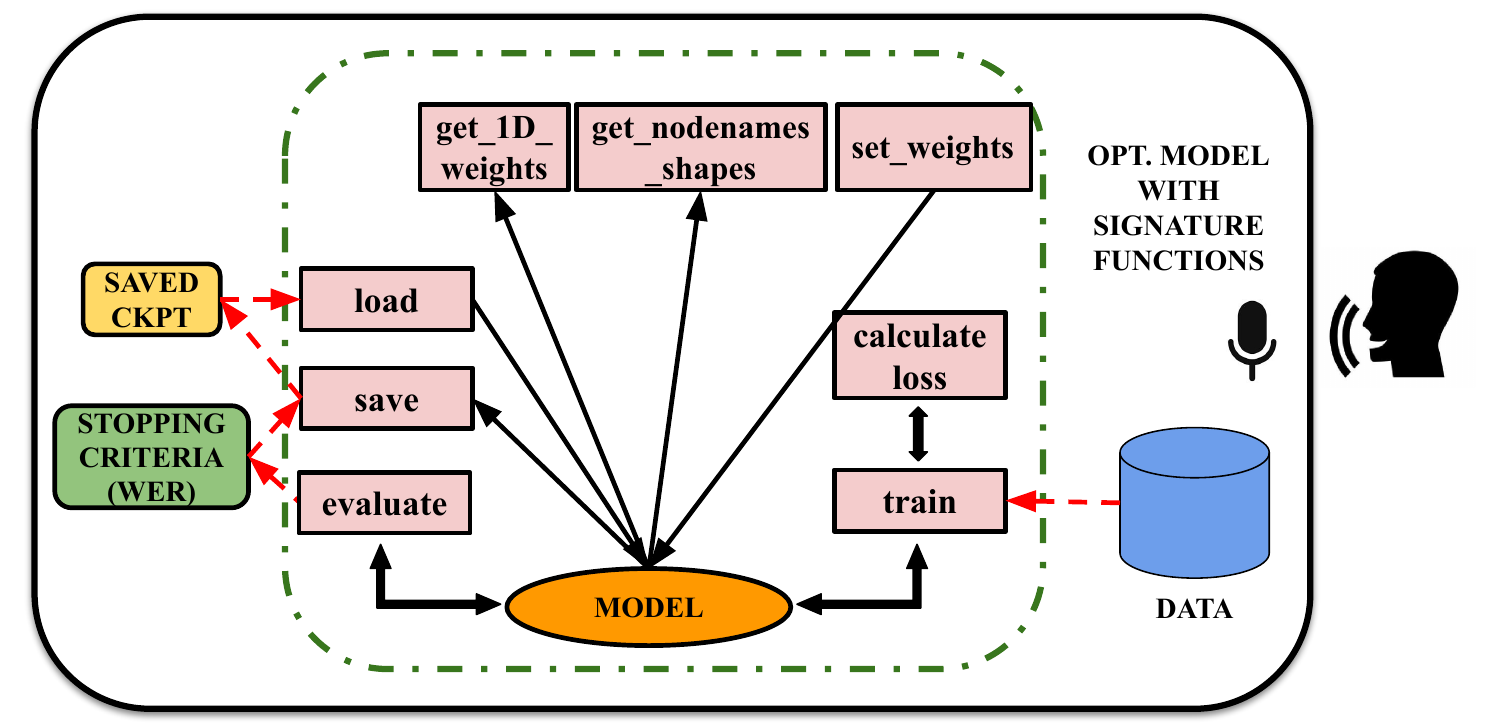}
	\caption{Optimised model created for on-device training and FL setup}
	\label{fig:tflite}
    \vspace{-0.5cm}

\end{figure}

It is critical to develop a better methodology for converting a larger global model to a memory efficient and optimised version suitable for edge devices. We use TensorFlow \cite{tf} to create a well-optimised Flatbuffer format based conversion, which allows for significant model size reduction. We use the signature functions in Tensorflow to interface with the optimised model and perform operations such as training, inference, loading and saving the checkpoint weights, as described in \cite{tflite,ourarxiv}. Furthermore, we build extra signature functions specifically for FL configuration to aid in efficient communication between clients and server. Figure \ref{fig:tflite} depicts the optimised model with eight signature functions that allow us to successfully train the model on the device, and perform FL related functions in the mobile devices.
Along with the existing signature functions such as train, evaluate, save, load, and calculate\_loss, we build three new signature functions: 
\begin{itemize}
	\item Get\_1D\_weights: reshapes an N-dimensional weight tensor from each node in the model graph to a 1-D array and returns a list of 1-D arrays.
	\item Get\_nodenames\_shapes: returns all the node names as well as the actual tensor shapes.
	\item Set\_weights: reshape the aggregated 1-D weight array to N-D tensor and reloads it into the model.
\end{itemize}

%Figure \ref{fig:tflite} depicts the optimised model with eight signature functions that allow us to successfully train the model on the device, and perform FL related functions in the mobile devices. 

One key functionality of these signatures is packing and unpacking the weights into 1-D and N-D arrays respectively. In \cite{credit}, the authors explained that important sensitive data can be decoded if an intruder gets access to weights during communication. But if the weights are packed into a 1-D array, private information like shape of the weight matrices of each layer which can give useful insights about the model architecture and data patterns are hidden. These three main signatures, Get\_1D\_weights, Get\_nodenames\_shapes and Set\_weights not only facilitate in implementing Ed-Fed framework but also induce privacy to the framework.  
% !!
% During training, we use a stopping criteria based on WER to prevent the model weights from degrading due to continuous training. When we see no improvement in WER for the first time, we train for two more epochs and halt the training if WER fails to show any further improvement.

Due to the limited resources available in the edge devices, we prevent repeated storing of checkpoints after each epoch of training.  Instead, we update a single checkpoint throughout the training. Hence, the optimised model (tflite) generated with the signature functions has all the functionalities which include support for training the models efficiently on the device as well as FL-based weight updation. The optimised model is utilised in our android application and our simulation systems, hence facilitating end-to-end real-time FL.
 \subsection{Communication protocol}
\label{sec:grpc}

We use the gRPC \cite{grpc} communication protocol to ensure efficient communication between the server and clients. gRPC, like many RPC systems, is built on the concept of establishing a service, which describes the methods that can be called remotely with their input and return types. Protocol buffers are the most common interface definition language (IDL) used by gRPC to describe both the service interface and the message structure of the payloads.
In our framework, three RPCs were used. ``CommunicatedText" is the first, which is used to send the current context of the client and server. ``GetGlobalWeights" is the second, which is used to send the current global FL weights to the clients who have been chosen for training. The final one is ``GetFLWeights," which is used to share the aggregated weights between the server and client.
\subsection{Server}
\label{sec:server}
On the server side, there are two major components: the \textit{Client selection with fairness} and \textit{Aggregation strategy}. 
Client selection involves selecting $k$ clients out of $N$ available clients. In case of waiting time optimised client selection algorithms, the clients should be selected in such a way that the overall waiting time of the clients is minimised, while ensuring fairness in the selection of clients. More about the client selection algorithms is discussed in further sections. 
% Training time varies device-to-device based on the resources available and computing power. So, if a device with weaker computing power and resources ($D_1$) and a device with abundant computing power and resources ($D_2$) are selected in the same FL round, $D_2$ will complete its training long before $D_1$ does and have to wait until $D_1$ completes its training. This is where client selection algorithms come into the picture and takes care of these abnormalities by minimising the waiting time of each client. However, there is a chance that this client selection algorithm might get biased towards stronger clients (more resources and computing power). So, the server should select clients such that the waiting time is minimised while maintaining fairness in the selection of clients.

The weight aggregation is an integral part of FL. The server strategy algorithms aggregates the weights ($w_t^i$) obtained from the selected $k$ clients by choosing algorithms such as FedAvg \cite{FL}, FedProx \cite{fedprox}, etc., and the updated weights are sent back to the clients. Typically, in real-world settings the client's local data will not be representative of the global data distribution because of the noisy background or the different voice characteristics. Hence, simply aggregating weights from such clients with low quality data will lead to a deviation from the global model. An improved solution is to generate a weighted aggregation of the models, with a weighting coefficient reflecting the quality of the model \cite{flasr_wer}. A client with a higher word error rate (WER) denotes poor performance of the global model. In such cases, we assign a lower weighting coefficient to that model during the aggregation. We use a weighted WER-based strategy in our Ed-Fed setting and is denoted as follows:
\vspace{-0.34cm}
\begin{align}
\label{eq:Strategy1}
 &w_{t+1} \gets \sum_{i=1}^k\alpha_iw^i_{t+1}\\
 \label{eq:Strategy2}
 &\alpha_i = \frac{\exp( 1- {WER}_i)}{\sum_{j=1}^k \exp( 1 - {WER}_j)}
\end{align}
where the weights $\alpha_i$'s are calculated using the softmax distribution obtained from the WER values from multiple clients. 

Once clients are selected for training, the server notifies them to start training. The clients begin their training, and once they complete their training, they send their updated weight to the server. The server then aggregates them through a weight aggregation strategy. The aggregation strategy considers the uncertainties in the quality and quantity of data by using the WER based weighting used by each client obtained during the training process.

\section{Resource-aware time-optimised client selection}
\label{cmab_basics}

\subsection{Need for waiting time optimisation}
\label{sec:need_for_waiting_time}
\begin{figure}
\begin{minipage}[t]{0.475\columnwidth}
  \includegraphics[width=\linewidth]{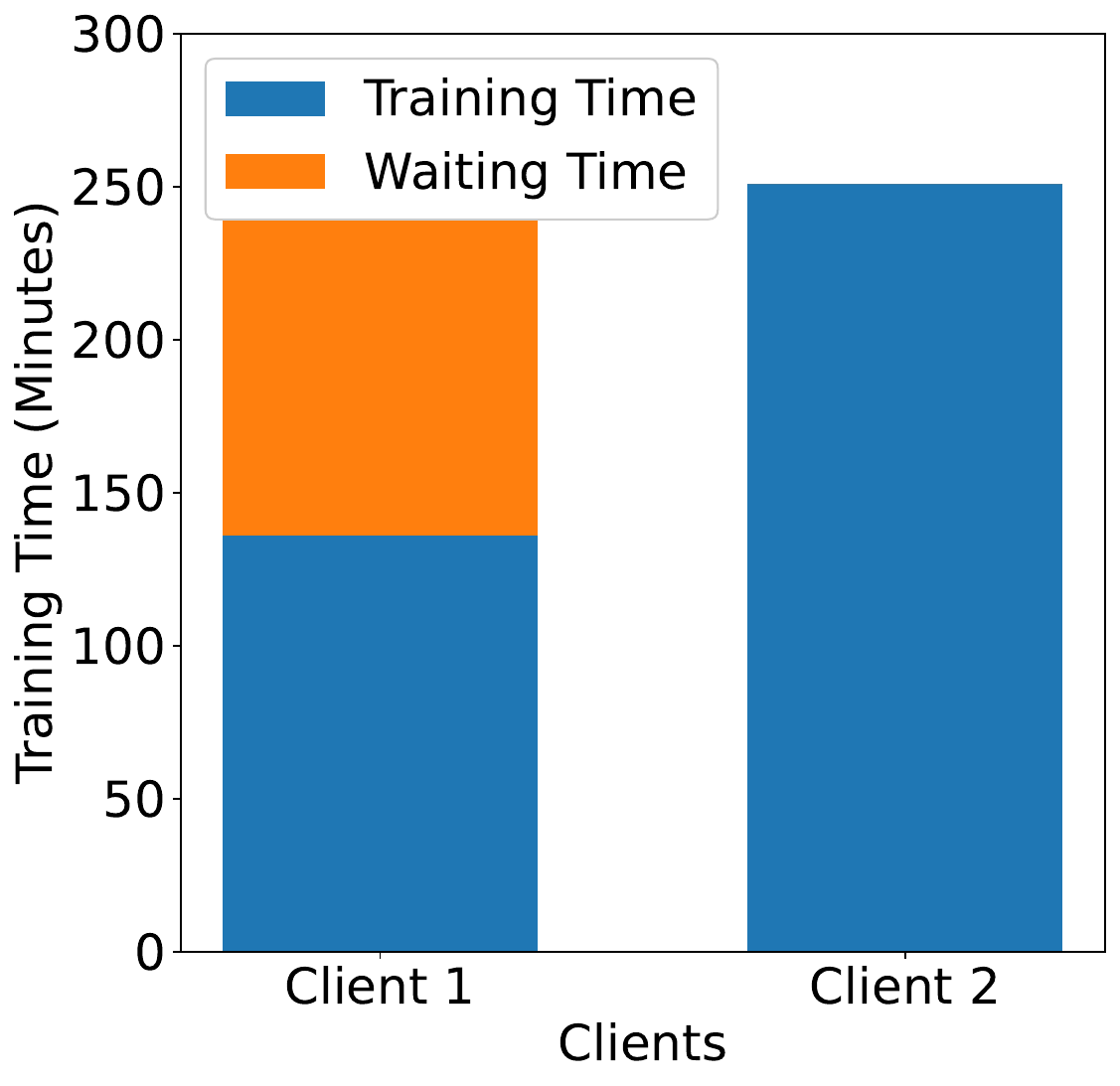}
    \vspace{-0.5cm}
  \caption{Scenario 1: Slow vs Fast client.}
  \label{fig:random_waiting_time1}
\end{minipage}
\begin{minipage}[t]{0.475\columnwidth}
  \includegraphics[width=\linewidth]{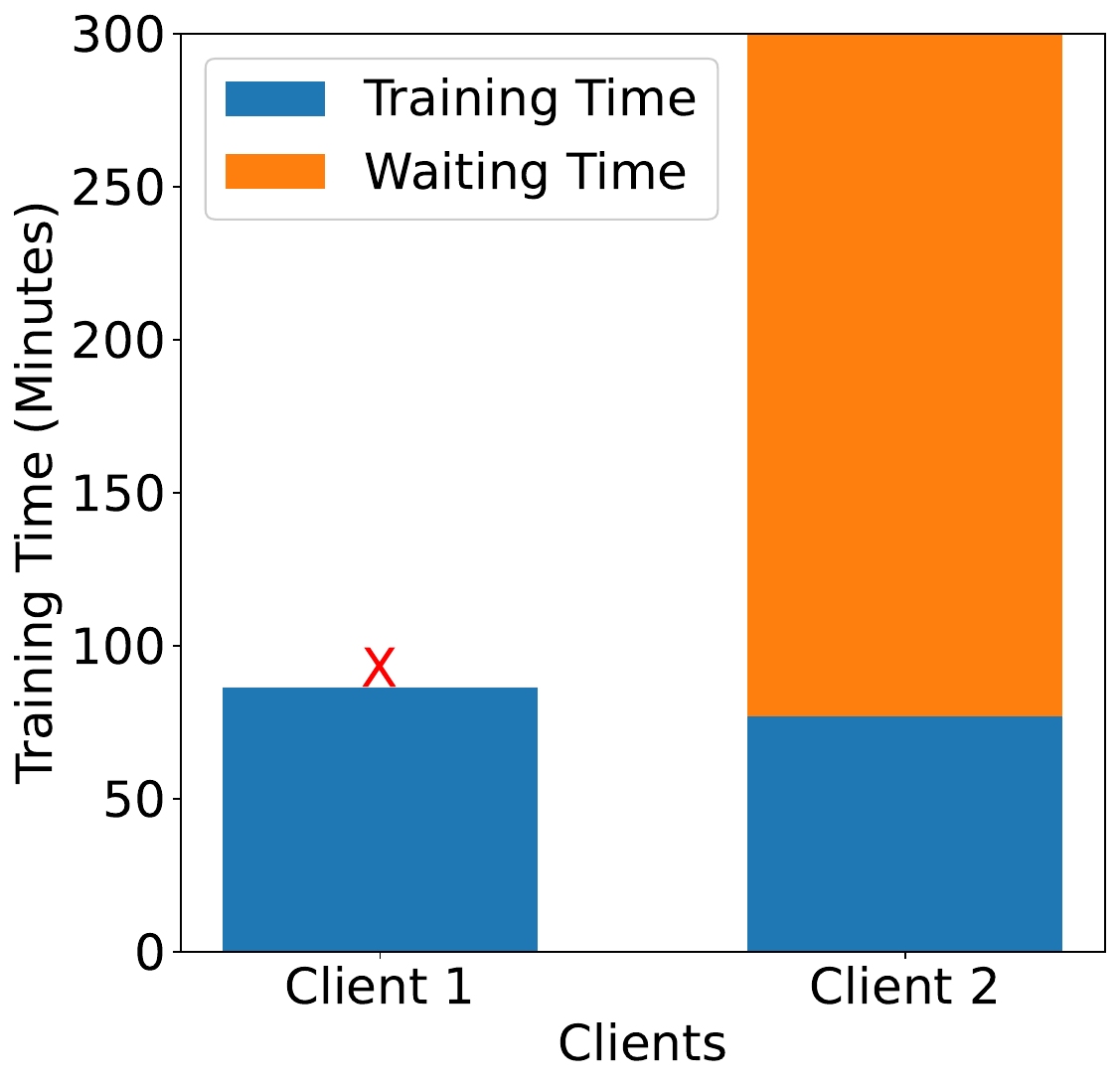}
   \vspace{-0.5cm}
  \caption{Scenario 2: One client with insufficient battery life.}
  \label{fig:random_waiting_time2}
\end{minipage}%\hfill% maximize horizontal separation
\end{figure}

% Figure \ref{fig:random_waiting_time1} depicts the training time and waiting time that are obtained during different experiments. 

Waiting time is the amount of time the client waits for server to fulfill the request. This is mainly observed when the clients send their updated weights and request the aggregated weights to be returned by the server. Since client platforms have different training times, the server cannot compute the aggregated weights until responses are obtained from all clients, leading to higher waiting times for clients with faster computing capabilities. 

To quantify the waiting time, we considered two common scenarios : (1) Select one fast client and one slow client and (2) Select one client with insufficient battery life made to run more number of epochs. Figure \ref{fig:random_waiting_time1} shows client 1 having higher compute capabilities which has waited for a significant amount of time for the slower client 2 to finish its training process.  Figure \ref{fig:random_waiting_time2} presents the results of the experiment (Scenario 2). We can see that client 1 switches off in the middle of training,  making client 2 wait for an infinite amount of time.

Considering all these problems, we created a resource-aware time-optimized client selection algorithm which takes the resources of the clients into consideration and assigns work adaptively to each client depending on that client's resources.
Our approach has three important phases, a resource information extraction, neural reward generation and a resource-aware time-optimised algorithm for clients.

% Initially, the server is blind to the resources available in the clients and hence exploration is important to ascertain these utilities available. It is also critical to use the information gathered from prior participations and exploit the clients with better resources. Our idea is to strike a balance between exploration and exploitation among the clients and adapt CC-MAB for a better client selection using the upper confidence bound (UCB) policy.
\subsection{Resource information}
\label{resources}
 We aim to investigate the relationship between the resource availability and the fluctuations in the training time. We also aim to assess the impact of training on  battery drain, which is crucial in real-world scenarios to prevent device shutdown during extended usage. The different resources which we considered for estimating the training time and battery drain of a particular phone are (1) memory-related information, (2) power-related information, (3) CPU-usage-related information (CI), and, (4) phone-specific information (PI). Memory-related information is captured using: (a) total RAM ($TR$), and, (b) available RAM ($AR$). In case of power-related information, (a) available battery charge ($AC$) and (b) battery status ($BS$) are used. For $CI$, the average usage of CPU across multiple cores is used. In order to find the $PI$, we used the Antutu score. The breakup of these resources into individual components is important because each of these can individually effect the training time. %Section \ref{sec:resources_effect} explains in detail how each resource effects the phone's training time. 
All these resources together form the context information of the client device, i.e., $c=\left[TR,AR,AC,BS,CI,PI\right]$. Obtaining context information from the phone before each FL round is important as  time taken to complete the training process depends on it. 
 We make use of this information by training a neural network to understand these dependencies and predict the expected training time and battery charge consumption based on a context vector. Once, we get to estimate these two parameters of each client, we can group clients together in a manner that minimises the waiting time of each client.

\subsection{Neural reward generator}
\label{sec:neuralucb}

\RestyleAlgo{ruled}
\begin{algorithm}[!t]
\caption{NeuralUCB-m}\label{alg:neuralucb}
\KwData{$\text{Number of training rounds}~T, \text{exploration parameters}$\\
$~\{\alpha_t\}_{t\in T},\text{Hidden layer size}~m ,\text{Context vectors at }t^{th}$\\
$ \text{ round: } \{c_{t,s}|s\in\{1\cdots N\}\}$
}
$\text{\textbf{Initialisation:}}~\text{Randomly initialise}~\boldsymbol{\theta_{0}^s},$
$\mathbf{Z}_{0,s}=\lambda\mathbf{I}~ \forall s\in {1 \cdots N}$\\
\For{t=1 \emph{ \KwTo} T}{

\For{s=1 \emph{ \KwTo} N}{
$\left[\hat{b\_t}_{t,s}, \hat{d}_{t,s}\right] \gets f_s(\mathbf{c_{t,s}};\boldsymbol{\theta}_{t-1}^s)$\\
$U_{t,s} \gets -\hat{b\_t}_{t,s} + \alpha_t \sqrt{\nabla{f_s(\mathbf{c_{t,s}};\boldsymbol{\theta}_{t-1}^s)}^T\mathbf{Z}_{t-1,s}^{-1} \nabla{f_s(\mathbf{c_{t,s}};\boldsymbol{\theta}_{t-1}^s)}/m } $
}
$S_t \gets \text{Top}~min(k, |U_t|)~\text{clients from}~U_{t}$\\
$\text{Run the clients}~s\in S_t~\text{and get the true } $$\left[{b\_t}_{t,s},{d}_{t,s}\right]$\\
$\text{Update}~\mathbf{Z}_{t,s} \gets \mathbf{Z}_{t-1,s} + \nabla{f_s(\mathbf{c_{t,s}};\boldsymbol{\theta}_{t-1}^s)}^T\nabla{f_s(\mathbf{c_{t,s}};\boldsymbol{\theta}_{t-1}^s)} /m$\\

$\mathbf{D}_{t,s} \gets \mathbf{D}_{t-1,s} \cup (\mathbf{c_{t,s}},\left[{b\_t}_{t,s}, {d}_{t,s}\right])$\\

$\boldsymbol{\theta}_{t}^s \gets \text{TrainNN}\left(\mathbf{D}_{t,s}, \boldsymbol{\theta}_{t-1}^s \right)$\\

}
% $No of epochs each client i should run \gets floor(n_{t,i}/(m_t/(\hat{b\_t}_{t,i})))$\;
\end{algorithm}
In this section, our goal is to learn the training time, and battery drop based on the given resources. Further, use these values to learn a better client selection by removing straggler clients or by doing an adaptive setting of the training time for clients. We will first give a brief overview of the basic contextual combinatorial multi-armed bandits (CC-MAB) problem using neural upper confidence bound (NeuralUCB)\cite{neuralucb} policy. We will then  use this in our FL framework setting to optimise the overall waiting time, based on the context information.

We formulate our client selection problem as a CC-MAB setting with $N$ arms in which the agent interacts with the arms for $T$ rounds. Initially, the true rewards generated by the arms are unknown, and the agent observes only the $N$ context vectors from corresponding arms: $\{\mathbf{c_{t,s}} \in \mathbb{R}^{d}|s\in\{1\cdots N\}\}$. Let $ \mathbf{S} \subseteq 2^{N} $ be the set of all proper subsets of $N$ available arms in the setting. At each round $t$, the agent predicts the rewards $\{\hat{r}_{t,s}\}_{s\in \mathbf{S_t}}$ using a reward estimating function $(f)$ parameterised by $\boldsymbol{\theta}$, such that  $\hat{r}_{t,s} = f(\mathbf{c_{t,s}};\boldsymbol{\theta}) $. Then, the agent selects a subset, $\mathbf{S_t} \in \mathbf{S}$ containing $k$ arms based on the calculated rewards.  In most cases, the agent is aware of the parametric form of the reward estimating function that is being used. In linear upper confidence bound problems (LinUCB)\cite{linucb}, for example, the reward is calculated as $\hat{r}_{t,s} = \boldsymbol{\theta_{*}}^T\mathbf{c_{t,s}}$. After playing the arms in $\mathbf{S_t}$, the agent observes the true rewards given as $\{r_{t,s}\}_{s\in \mathbf{S_t}} $.
The goal of the agent is to maximise the expected reward, which is equivalent to minimising $R_T$, the cumulative regret over $T$ rounds: 
\begin{equation}
    R_T = \mathbb{E} \left[\sum_{t=1}^{T} \left( \sum_{s \in \mathbf{S^{*}_t}}r_{t,s} -  \sum_{s \in \mathbf{S_t}}\hat{r}_{t,s}\right)             \right] 
\end{equation}
where $\mathbf{S^{*}_t}$ is the set of $k$ optimal arms with maximum true rewards at round $t$. 

Unlike the classical linear contextual bandit where the reward estimating function $f(\mathbf{c_{t,s}};\boldsymbol{\theta})$ is linear, we utilises a neural network to deal with the intricate relationship between context features and rewards. The neural network based reward estimating function $f(\mathbf{c_{t,s}};\boldsymbol{\theta})$
estimates the expected reward of an action based on past observations, and is parameterised as:
\begin{equation}
f(\mathbf{c_{t,s}};\boldsymbol{\theta}) = \sqrt{m}\mathbf{W_L}\sigma\left(\mathbf{W_{L-1}}\sigma\left(\cdots \sigma\left( \mathbf{W_1}\left( \mathbf{c_{t,s}} \right)\right)\right)\right)
\end{equation}
where $L$ is the total number of hidden layers, $m$ is the hidden layer size (assumed to be same for all the layers for convenience), $\sigma$ is the activation function, and $\mathbf{W_l}$ corresponds to the weight matrix in $l^{\text{th}}$ layer. Here $\boldsymbol{\theta}$ is the vectorised weight matrices from all the hidden layers in the neural network and is given as $\boldsymbol{\theta} = \left[\text{vec}(\mathbf{W_1}^{T}),\text{vec}(\mathbf{W_2}^{T}) \cdots \text{vec}(\mathbf{W_L}^{T})   \right]$. 

Our proposed approach for client selection considers the clients or edge devices participating in FL as the arms with resource information discussed in Section \ref{resources} as the context vectors. 
 We also observe that using a single reward generation model $f(\mathbf{c_{t,s}};\boldsymbol{\theta}_{t-1})$ for multiple client devices (NeuralUCB-s) may lead to performance degradation if the edge devices have different intrinsic characteristics such as age and usage history. For example, consider two identical phones, one of which has been in use for 5 years and the other of which is brand new. However, if we calculate the training time and battery drop of both phones under similar contexts, they do not match. We see that the older phone performs badly and drains faster due to the aging of batteries. Also, it is dependent on how extensively the phone is used over time. If we use a single model, we miss out on these relationships, and the model gets confused if such clients exist in our rounds with degradation in performance. On the other hand, personalised reward generation models $f_s(\mathbf{c_{t,s}};\boldsymbol{\theta}_{t-1}^s)$ (NeuralUCB-m),  can better adapt to the unique characteristics of each client device and provide more accurate results. In our approach, the neural network predicts the training time per batch of samples and also the drop in battery percentage given the resource information as the context vector, i.e., $\left[\hat{b\_t}_{t,s}, \hat{d}_{t,s}\right]=f_s(\mathbf{c_{t,s}};\boldsymbol{\theta}_{t-1}^s)$. We use the negative of training time per batch ($-\hat{b\_t}_{t,s}$) for each arm as the reward in the UCB setting and the battery drop is used for straggler handling. Once the training is completed for round $t$, the clients will send their $\left[{b\_t}_{t,s}, {d}_{t,s}\right]$ along with the updated weights. Then, $(\mathbf{c_{t,s}},\left[{b\_t}_{t,s}, {d}_{t,s}\right])$ will be added to the corresponding clients' dataset ($\mathbf{D}_{t,s}$) and will be used for training the neural network. 
 
 The detailed algorithm for neural combinatorial contextual bandits is shown in Algorithm \ref{alg:neuralucb}.

% add about training loss (mse), and training details, why -time, b\_t (other notation)
% subfigure
% algorithm whats n?, n1,n2?

% \subsubsection{Server strategy}
% \label{sec:strategy}
% The weight aggregation is an integral part of FL algorithm. The server strategy algorithms aggregates the weights obtained from the selected $k$ clients by choosing algorithms such as FedAvg \cite{FL}, FedProx \cite{fedprox}, etc., and the updated weights are sent back to the clients. Typically, in real-world settings the client's local data will not be representative of the global data distribution because of the noisy background or the different voice characteristics. Hence, simply aggregating weights from such clients with low quality data will lead to a deviation from the global model. So, a better solution is to generate a weighted aggregation of the models, with a weighting coefficient reflecting the quality of the model \cite{flasr_wer}. A client with a higher WER denotes poor performance of the global model. In such cases, we give a lower weighting coefficient to that model during the aggregation. We use a weighted WER-based strategy  in our Ed-Fed setting and is denoted as follows:
% \begin{align}
% \label{eq:Strategy1}
%  &w_{t+1} \gets \sum_{i=1}^k\alpha_iw^i_{t+1}\\
%  \label{eq:Strategy2}
%  &\alpha_i = \frac{\exp( 1- {WER}_i)}{\sum_{j=1}^k \exp( 1 - {WER}_j)}
% \end{align}
% where the weights $\alpha_i$ are calculated using the softmax distribution obtained from the WER values from multiple clients. 

\begin{table*}[!t]
\vspace{0.3in}
	\centering
	\caption{Details about mobile phone hardware.}
	\label{tab:different_phones}
	\begin{tabular}[t]{lcccccc}
		\toprule
		Device Model&RAM&CPU&OS&SoC\\
		\midrule
		OnePlus 7T-1&4.0$|$8 GB&Octa-core Max 2.96 GHz&Oxygen 11 (Android 11)&Snapdragon 855 Plus\\
		OnePlus 7T-2&4.1$|$8 GB&Octa-core Max 2.96 GHz&Oxygen 11 (Android 11)&Snapdragon 855 Plus\\
		OnePlus 5T&3.6$|$6 GB&Octa-core Max 2.45 GHz&Oxygen 9 (Android 9 )&Snapdragon 835\\
        Xiaomi 11 Pro &4.8$|$8 GB&Octa-core Max 2.05 GHz&MIUI 13 (Android 11)&Helio G96\\
		\bottomrule
	\end{tabular}
 \vspace{-0.3cm}

\end{table*}%

\RestyleAlgo{ruled}

\begin{algorithm}[t!]
\caption{Resource-aware time-optimised algorithm for client selection}\label{alg:resource_aware}
\KwData{$\text{Context vectors of clients at } t^{th} \text{ round: } \{c_{t,s}|s\in\{1\cdots N\}\} $
$\text{Number of samples at round t: } (n_{t,1},n_{t,2}..,n_{t,N}),$
$\text{Number of clients to be selected: } k, \text{Minimum}$
$\text{ and maximum number of epochs to run in a FL} $
$\text{round: }(e_{min},e_{max}), \text{Batch size: } bs, $ Battery \text{ threshold} $\gamma.$
}
\KwResult{$\text{k clients along with their respective }e_{t,i}$}
% $j \gets 1$\;
% $Predict~\hat{b\_t}_{t,i},~\hat{d}_{t,i}~using~Neural~network~for~each~clientc_i~in~the~server$\;
% $\hat{b\_t}_{t,i} \gets time~taken~by~c_i~to~complete~one~batch~of~training$\;
% $\hat{d}_{t,i} \gets drop~in~the~battery~paercentage~of~c_i~for~one~batch~of~training$\;
% $Find~out~the~maximum~no~of~batches(b_{max_{t,i}})~each~client~can~run$\;

\For{i=1 \emph{\KwTo} N}{
% \tcc{\footnotesize{Step 1}}
$\hat{b\_t}_{t,i},~\hat{d}_{t,i} \gets f_i(\mathbf{c_{t,i}};\boldsymbol{\theta}_{t-1}^i)$ \tcc*[r]{\footnotesize{Step 1}} 
% \tcc{\footnotesize{Find the maximum number of batches and epochs each client can run}}
$b_{max_{t,i}} \gets \lfloor (AC_{t,i} - \gamma)/\hat{d}_{t,i} \rfloor $\;
$e_{max_{t,i}} \gets min(e_{max}, \lfloor (b_{max_{t,i}}/(n_{t,i}/bs)) \rfloor$) \tcc*[r]{\footnotesize{Step 2}}
% \tcc{\footnotesize{Filter out clients into M that can run at least $e_{min}$ epochs}}
\If(\tcc*[f]{\footnotesize{Step 3}}){$e_{max_{t,i}} \geq e_{min}$}{
$P_{t}.append(c_{t,i})$\;
}
}
% $Calculate the maximum no of epochs M_{e_{i}} each client can run from b_{max_{t,i}}$\;
% \For{i=0 \emph{\KwTo} n}{
% $M_{e_{i}} \gets floor(b_{max_{t,i}}/n_{t,i})$\;
% \If{$M_{e_{i}} \geq e_{min}$}{
% $U_i \gets UCB(c_i)$\;
% $M.append((U_i,~c_i,~i))\;$
% }
% }
% \tcc{\footnotesize{Pick the top k clients using UCB algorithm}}
$S_t \gets \text{NeuralUCB-m}(P_{t})$\tcc*[r]{\footnotesize{Step 4}}
% $M \gets \{c_i if e_{max_{t,i}} \geq e_{min}\}$\;
% $K \gets top k of UCB(M)$\;
% \tcc{\footnotesize{Find the maximum time the FL round can run}}
$m_t \gets 0$\;
\ForEach{client $i$ in $S_t$}{
$m_t \gets min(m_t,e_{max_{t,i}}*(n_{t,i}/bs)*\hat{b\_t}_{t,i})$ \tcc*[r]{\footnotesize{Step 5}}

% \eIf{$e_{max_{t,i}}>e_{max}$}{
% $m_t \gets min(m_t, e_{max}*(n_{t,i}/b)*\hat{b\_t}_{t,i})$\;
% }{
% $m_t \gets min(m_t, e_{max_{t,i}}*(n_{t,i}/b)*\hat{b\_t}_{t,i})$\;
% }
}
% $m_t \gets minimum(b_{max_{t,i}}*\hat{b\_t}_{t,i} for i in K)$\;
% \tcc{\footnotesize{Find the number of epochs each client should run}}
\ForEach{client $i$ in $S_t$}{
$e_{t,i} \gets \lfloor (m_t/\hat{b\_t}_{t,i})*(bs/n_{t,i}) \rfloor$ \tcc*[r]{\footnotesize{Step 6}}}

% $e_i \gets min(e_{max},~\lfloor (m_t/\hat{b\_t}_{t,i})*(bs/n_{t,i}) \rfloor$\;}
% $No of epochs each client i should run \gets floor(n_{t,i}/(m_t/(\hat{b\_t}_{t,i})))$\;
\end{algorithm}
\begin{figure*}
     \centering
     \begin{subfigure}[b]{0.24\textwidth}
         \centering
         \includegraphics[width=\textwidth]{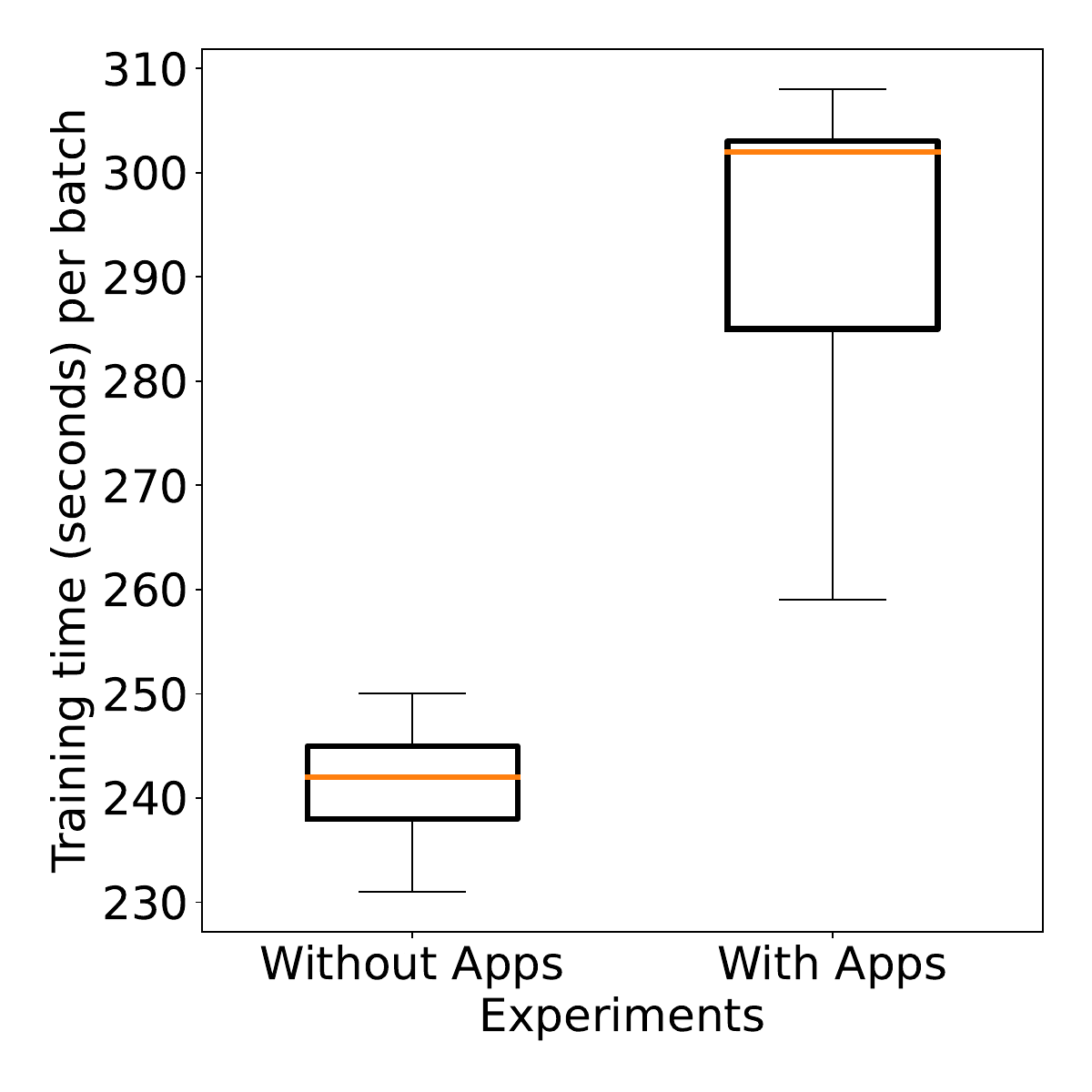}
         \caption{OnePlus 5T}
         \label{fig:1_ram}
     \end{subfigure}
     \hfill
     \begin{subfigure}[b]{0.24\textwidth}
         \centering
         \includegraphics[width=\textwidth]{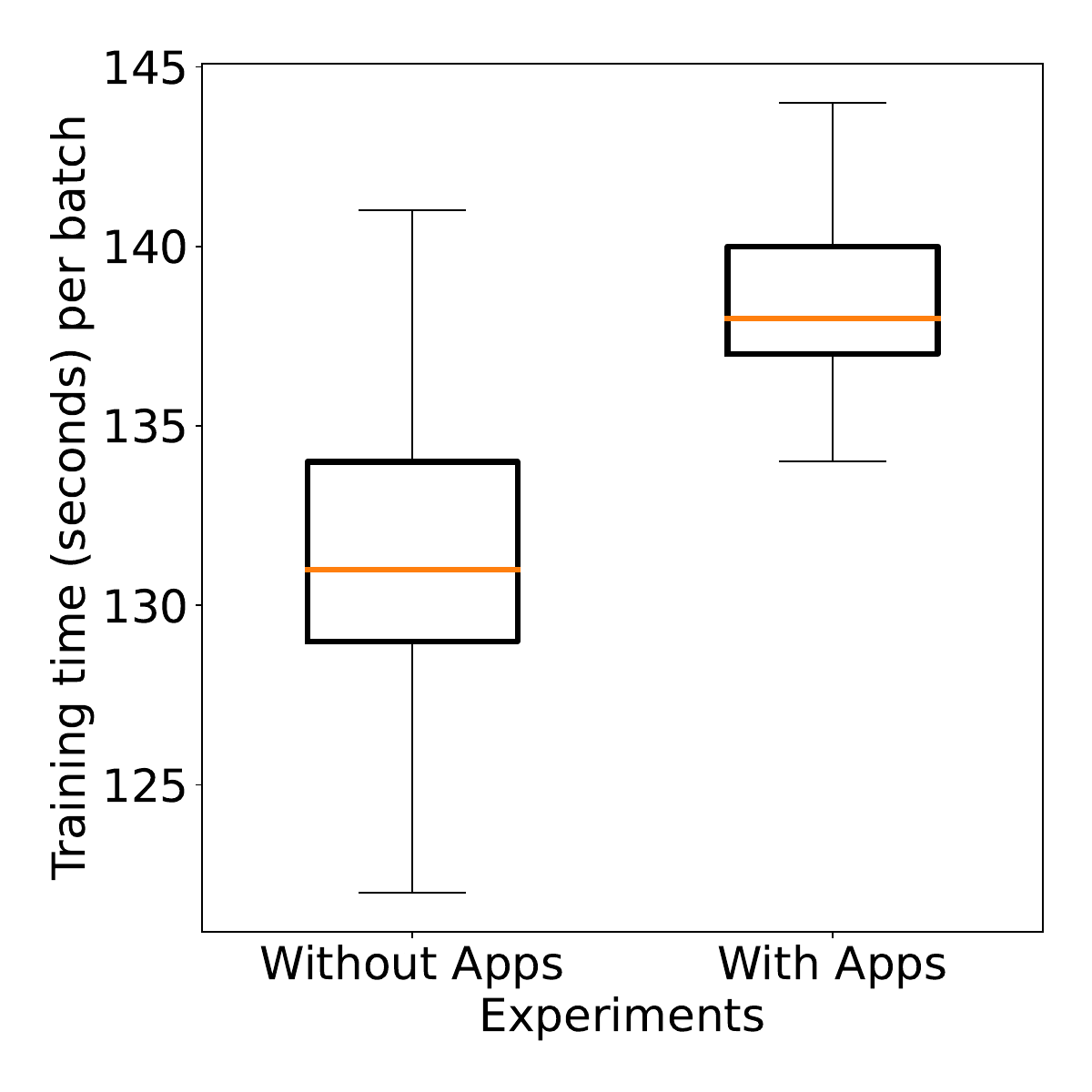}
         \caption{OnePlus 7T - 1}
         \label{fig:2_ram}
     \end{subfigure}
     \hfill
     \begin{subfigure}[b]{0.24\textwidth}
         \centering
         \includegraphics[width=\textwidth]{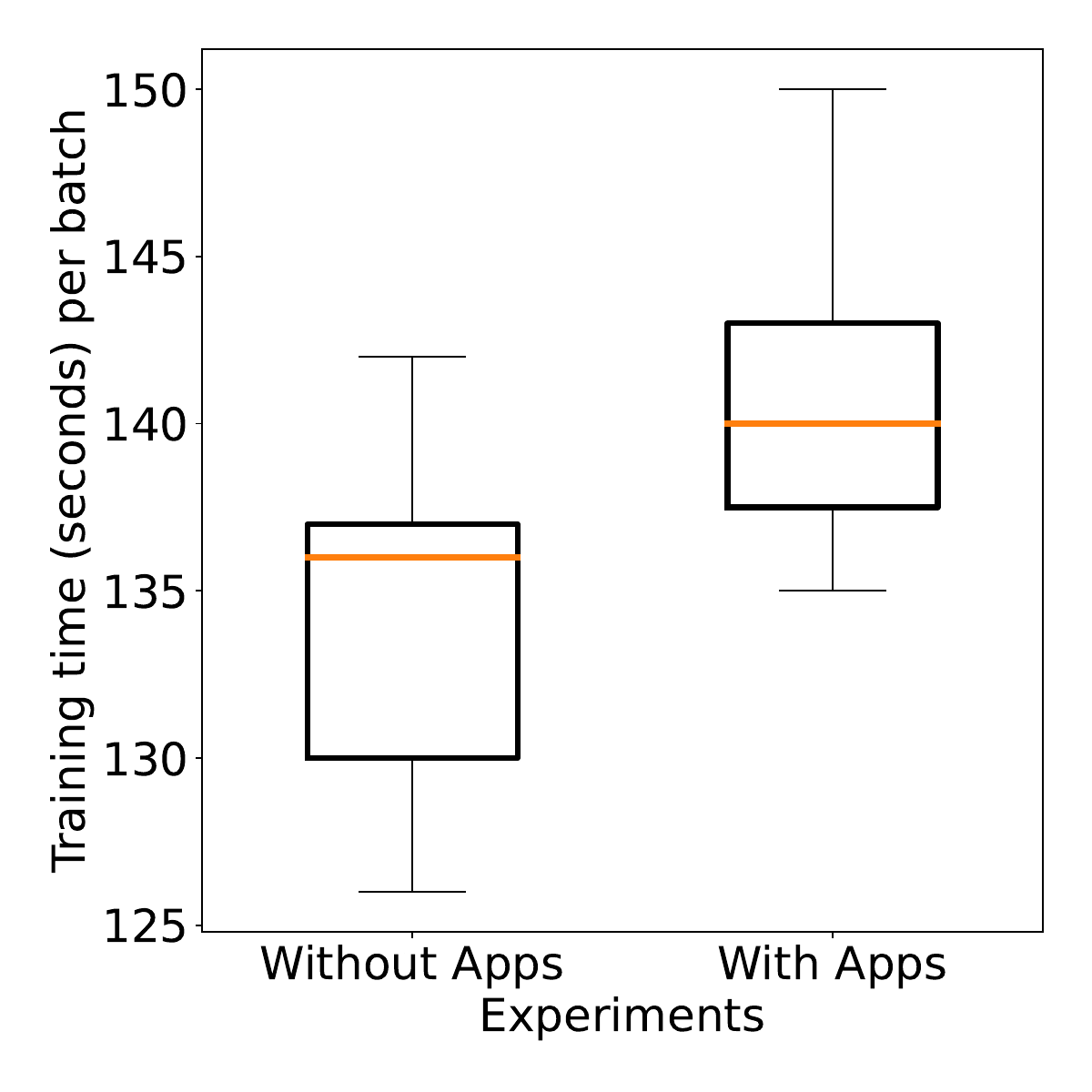}
         \caption{OnePlus 7T - 2}
         \label{fig:3_ram}
     \end{subfigure}
     \hfill
     \begin{subfigure}[b]{0.24\textwidth}
         \centering
         \includegraphics[width=\textwidth]{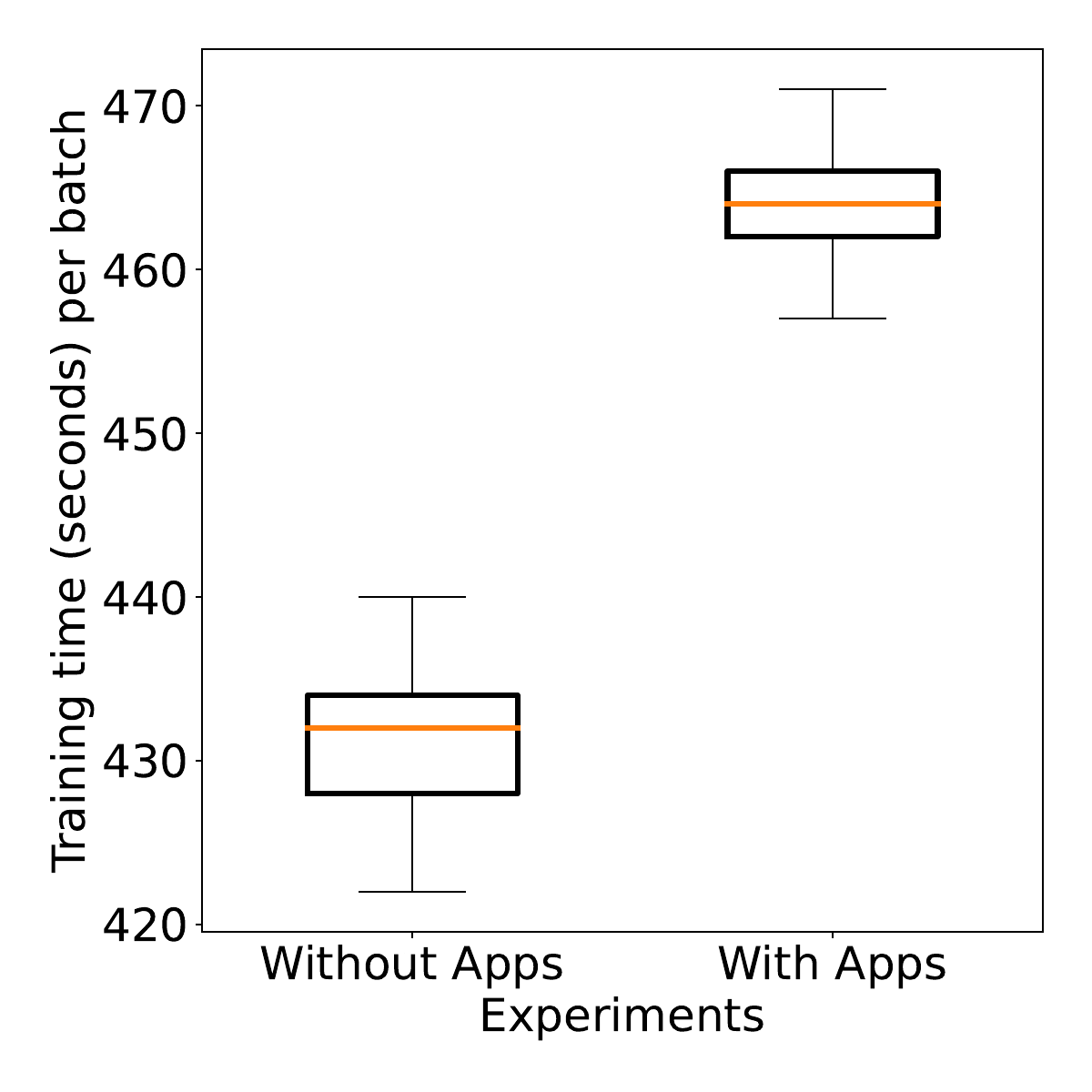}
         \caption{Xiaomi 11 Pro}
         \label{fig:4_ram}
     \end{subfigure}

        \caption{Effect of RAM on training time}
        \label{fig:ram_effect}

\end{figure*}
\subsection{Resource-aware time-optimised algorithm for client selection}

Initially, all the available clients $\{s\in\{1\cdots N\}\}$ who wish to participate in $t^{th}$  FL round, express their interest by sending their context vectors $\{c_{t,s}\}$ and number of samples available for training ($n_{t,i}$) to the server. The server then selects $k$ clients based on Algorithms \ref{alg:neuralucb} and \ref{alg:resource_aware} and notifies the selected clients about their selection along with the number of epochs ($e_{t,i}$) they need to run in that round which is computed using Algorithm \ref{alg:resource_aware}. The methodology is as follows:

\begin{enumerate}[Step 1]
    \item Calculate: (a) time taken to complete training a batch of data ($\hat{b\_t}_{t,i}$), (b) drop in battery on training a batch of data ($\hat{d}_{t,i}$), and (c) the maximum number of batches ($b_{max_{t,i}}$), each client can run while maintaining the battery above $\gamma\%$.
    \item Calculate the maximum number of epochs ($e_{max_{t,i}}$) each client can run from $b_{max_{t,i}}$, $n_{t,i}$, $e_{max}$ and $bs$
    \item  Filter out the clients who can run a minimum  of $e_{min}$ epochs into a set ($P_{t}$)
    \item Create a set ($S_t$) by picking $min(k, |P_{t}|)$ clients of $P_{t}$ using Algorithm \ref{alg:neuralucb}.
    \item Calculate the time taken to run $e_{max_{t,i}}$ epochs for each client and take the minimum of all these times ($m_t$). $m_t$ will be the maximum time the FL round can happen while ensuring that no client switches off in this time.
    \item Calculate the number of epochs ($e_{t,i}$) each client can run in $m_t$ time. 
    \item Notify each client of $S_t$ about their selection for that FL round along with their respective $e_{t,i}$.
\end{enumerate}

Thus, this algorithm ensures to adaptively choose the number of training epochs for the chosen clients while taking into account the available resources and minimising the waiting time for each client during FL rounds.

% !!
% In this section, we formulate the client selection algorithm for FL as a contextual combinatorial multi-armed bandit (CC-MAB) problem. The proposed algorithm intends to improve the efficacy of the client selection algorithm by considering the resource information from the clients, and thereby estimating the clients with ample computation resources to complete the training rounds. 

\section{Ed-Fed framework setup}
\label{sec:evaluation}
  We explain the entire setup used for evaluating Ed-Fed framework in both simulation and mobile environments.

\subsection{Setup for system-based evaluation}
In this section, we present the simulation settings used for evaluating our Ed-Fed framework for ASR tasks. The objective of this experiment to make the global model robust to multiple accents by learning from different accented clients. 

We use an end-to-end acoustic model similar to DeepSpeech2 \cite{ds2} architecture, collectively trained on datasets such as Librispeech \cite{librispeech}, commonvoice \cite{commonvoice} and tedlium \cite{tedlium} as our initial global model. For FL experiments, we created an audio corpus using a text-to-speech (TTS) system \cite{naturalreader} for 15 different accented speakers to simulate unique clients. Each speech sample is about 8-10 seconds, with an average label length limited to 150 characters. We run the FL experiments by associating one speaker data to one client. To evaluate the global model's accuracy, we created a test set consisting of speech samples from various speakers.

The simulation environment uses a single server and multiple python clients. We conduct a series of experiments, where we use our Ed-Fed framework to train the baseline ASR model on our audio corpus.  
We repeated the experiment for different values of k varying from 3 to 5. Each experiment is run for $T=5$ rounds with a fixed k and the k clients are randomly chosen from a pool of 10 readily available clients. We use the server strategy algorithm explained in Section \ref{sec:server} to aggregate the weights from the selected k clients. We train the model with 25 samples and a validation set of 10 samples.  All of our experiments were carried out using NVIDIA RTX 3090 and 3080 GPUs on a 10-core Intel i9-10900K CPU.

\subsection{Mobile phone based Evaluation}
With this set of experiments, we discuss the technical details associated with running Ed-Fed on edge devices. 
We present results from deploying our framework on our custom-built android application which allows the user to record speech samples. These recorded samples gets stored into the local memory of the application. 

We save the datasets for training and testing in the storage cache of the mobile phone. 
% The application supports the user to either record the samples or use the dataset available in the cache. 
We train the model with 25 samples and a validation set of 10 samples.
We host our python Ed-Fed server with WER based aggregation strategy on a local machine. The Ed-Fed clients are the Android mobile phones listed in Table \ref{tab:different_phones}. We use the optimised model mentioned in Section \ref{sec:clients} for the on-device personalisation of the ASR model.

\section{Results}
\label{sec:expsettings}

 \begin{figure}[!t]
    \centering
	\includegraphics[width = 0.35\textwidth,height=5cm]{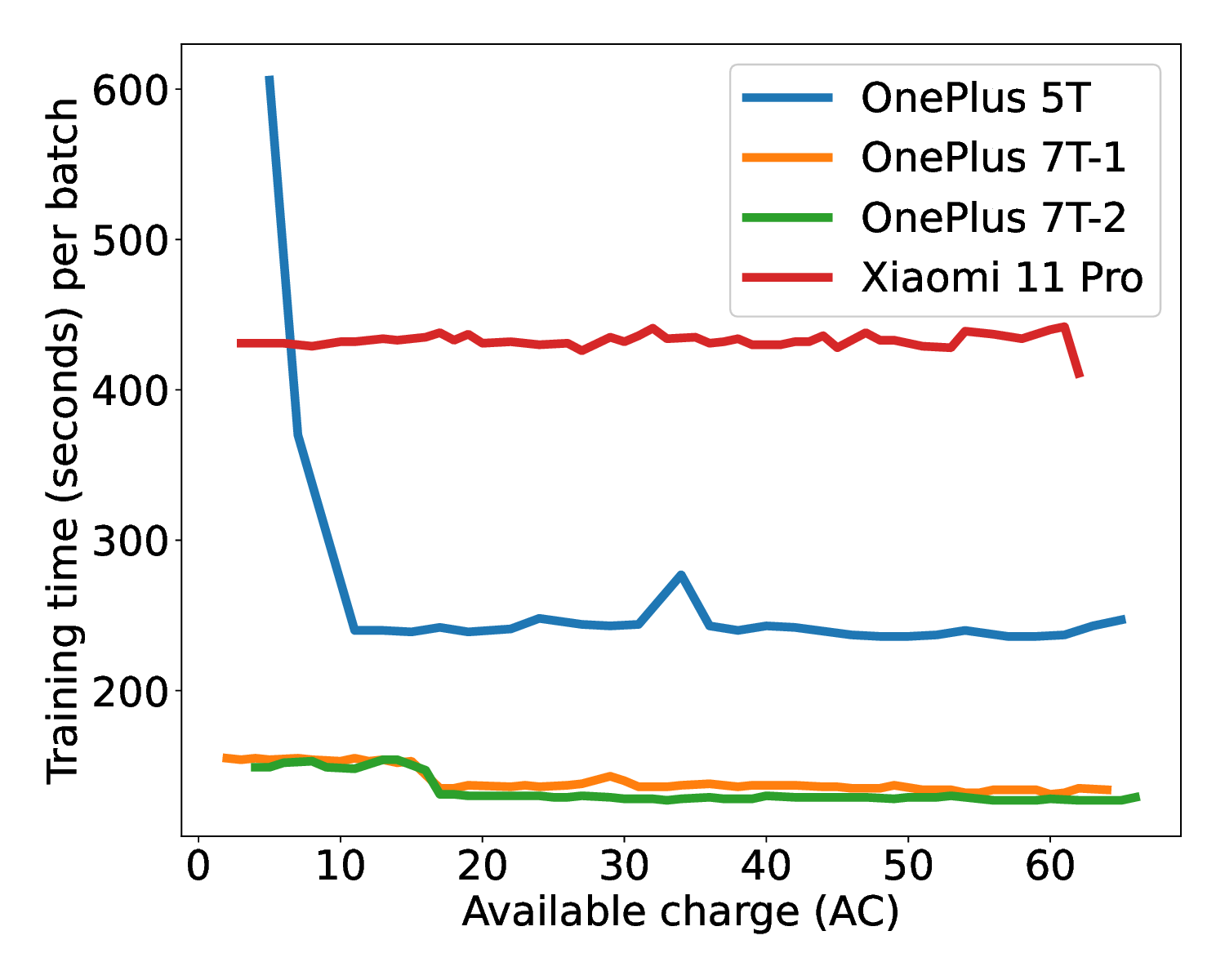}
             \vspace{-0.35cm}
	\caption{Battery vs training time}
	\label{fig:battery_effect}
 \vspace{-0.25cm}

\end{figure}

\begin{figure*}
     \centering
     \begin{subfigure}[b]{0.3\textwidth}
         \centering
         \includegraphics[width=\textwidth,height = 4cm]{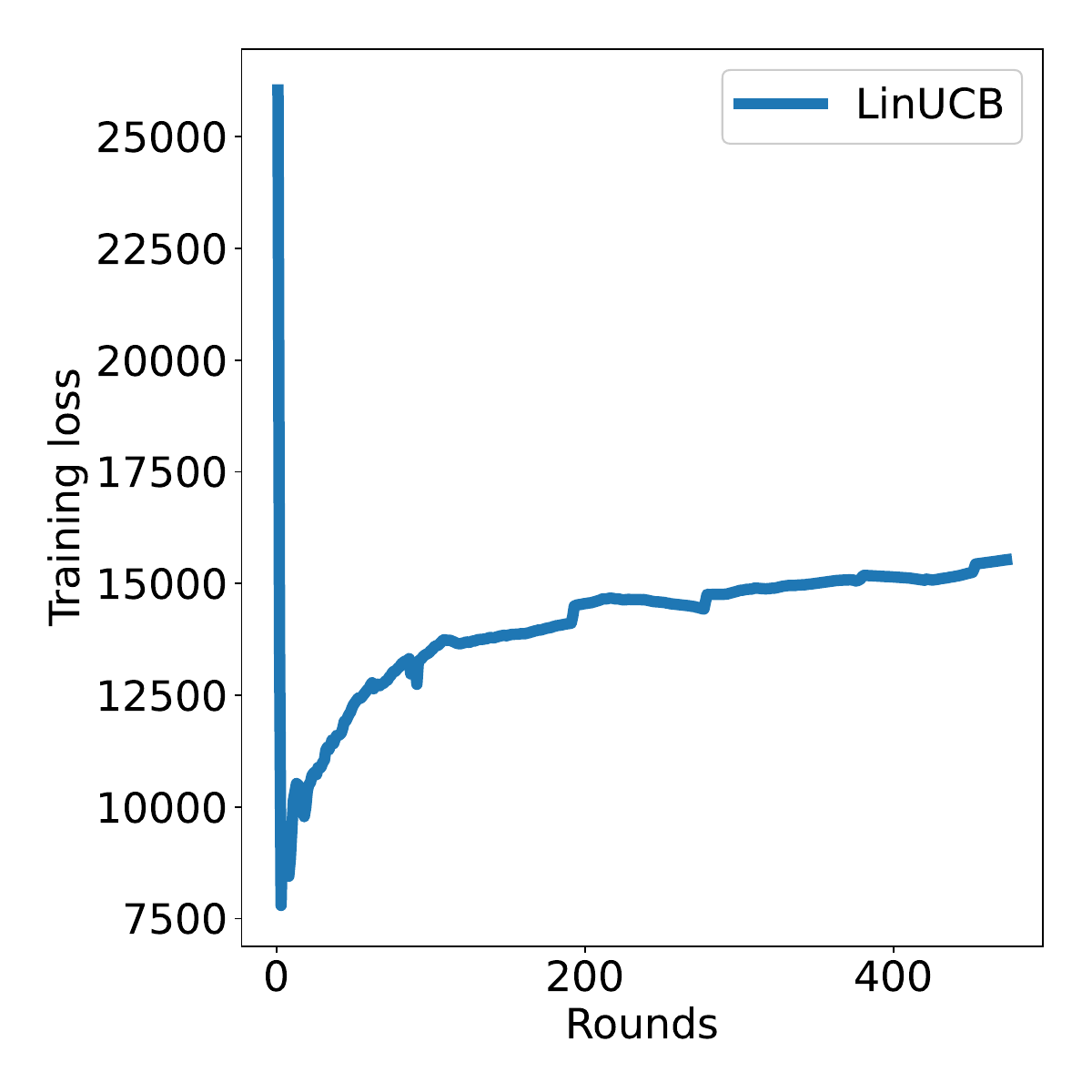}
            \vspace{-0.5cm}
         \caption{LinUCB}
         \label{fig:linucbmse}
     \end{subfigure}
     \hfill
     \begin{subfigure}[b]{0.3\textwidth}
         \centering
         \includegraphics[width=\textwidth,height = 4cm]{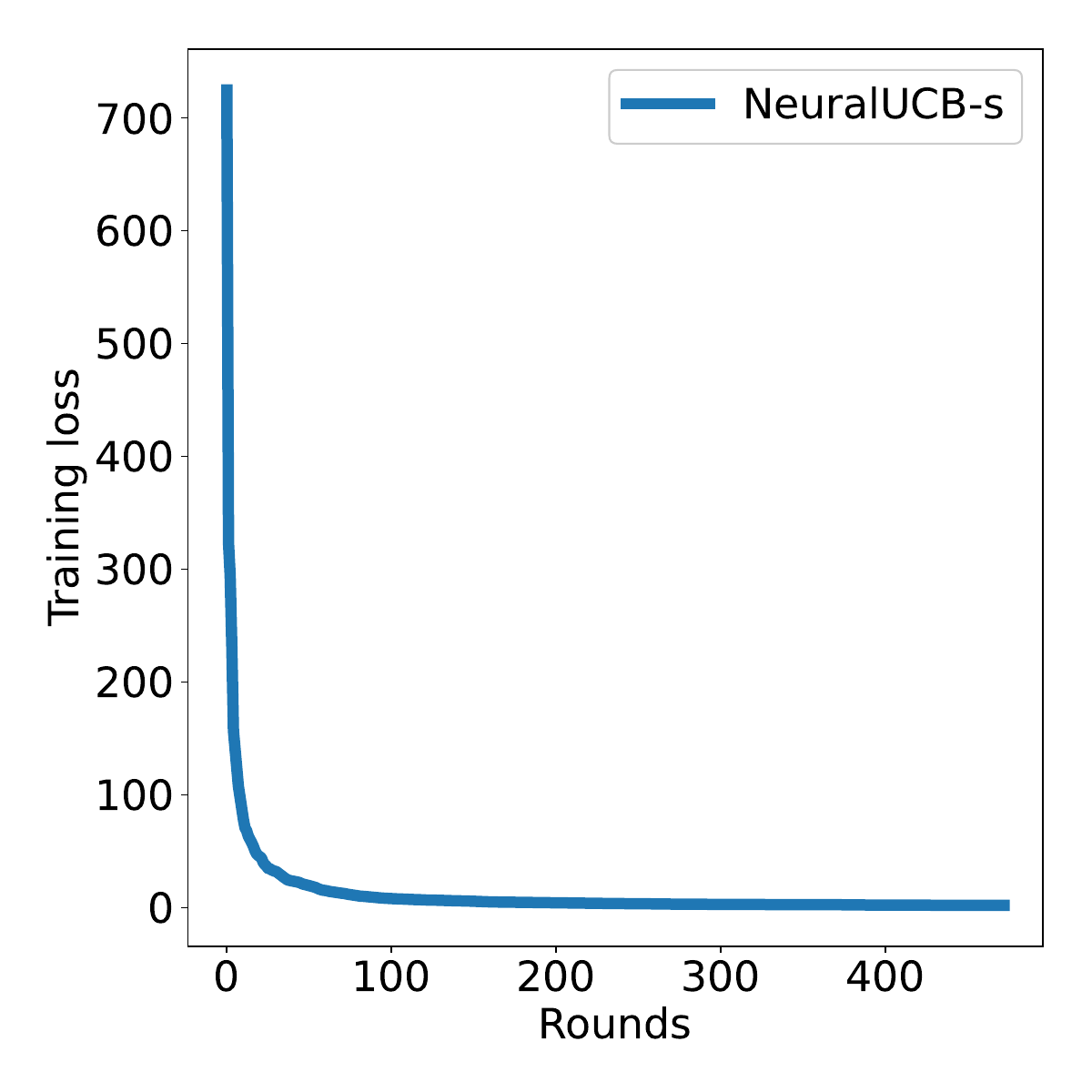}
            \vspace{-0.5cm}
         \caption{NeuralUCB single model}
         \label{fig:neuralucb-s}
     \end{subfigure}
     \hfill
     \begin{subfigure}[b]{0.3\textwidth}
         \centering
         \includegraphics[width=\textwidth,height = 4cm]{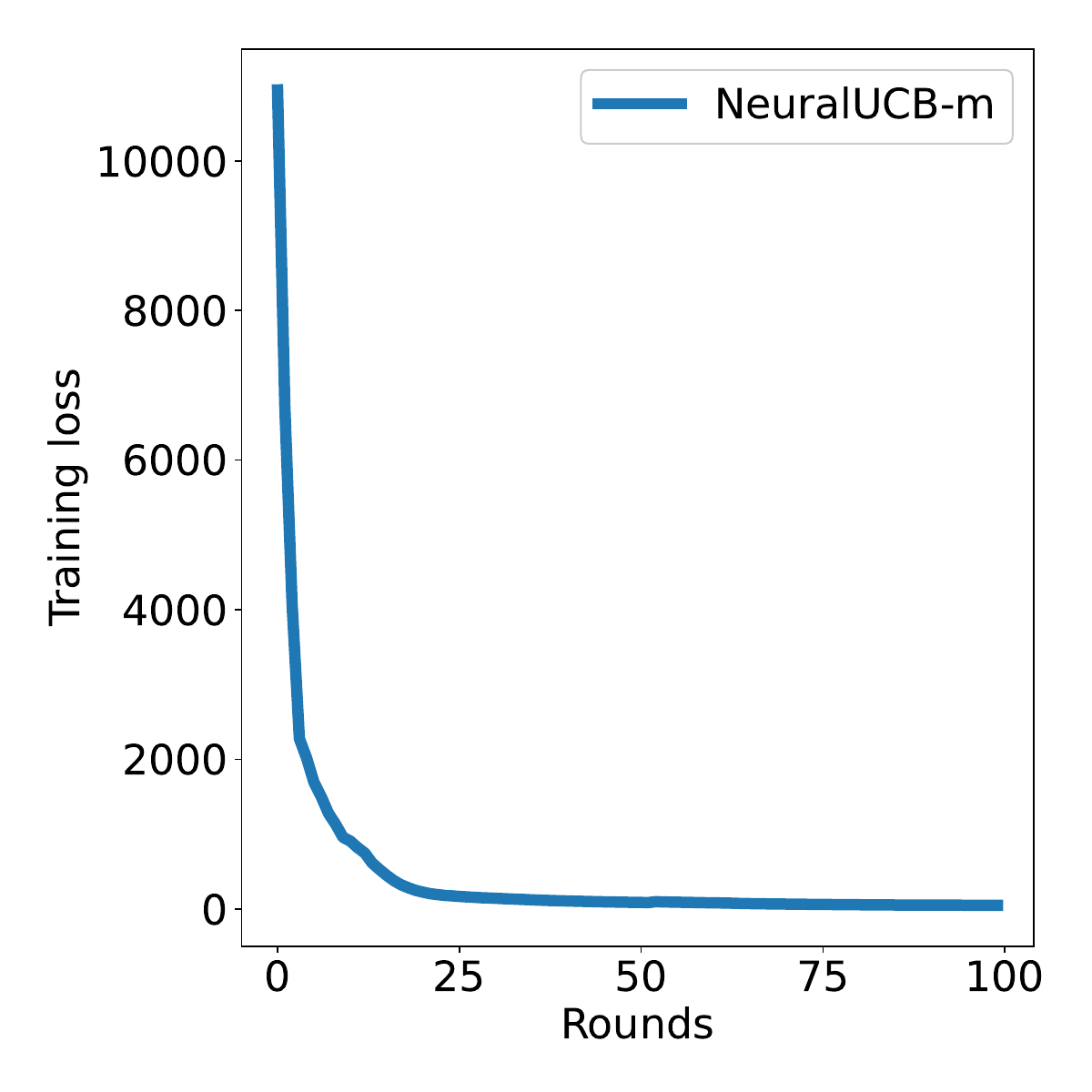}
            \vspace{-0.5cm}
         \caption{NeuralUCB multiple models}
         \label{fig:neuralucb-m}
     \end{subfigure}
        \vspace{-0.2cm}
     \caption{Training loss vs FL rounds for different reward generator functions}
     \label{fig:mse_plots}
\end{figure*}

\begin{figure}[!t]
    \centering
	\includegraphics[width = 0.35\textwidth,height = 5cm]{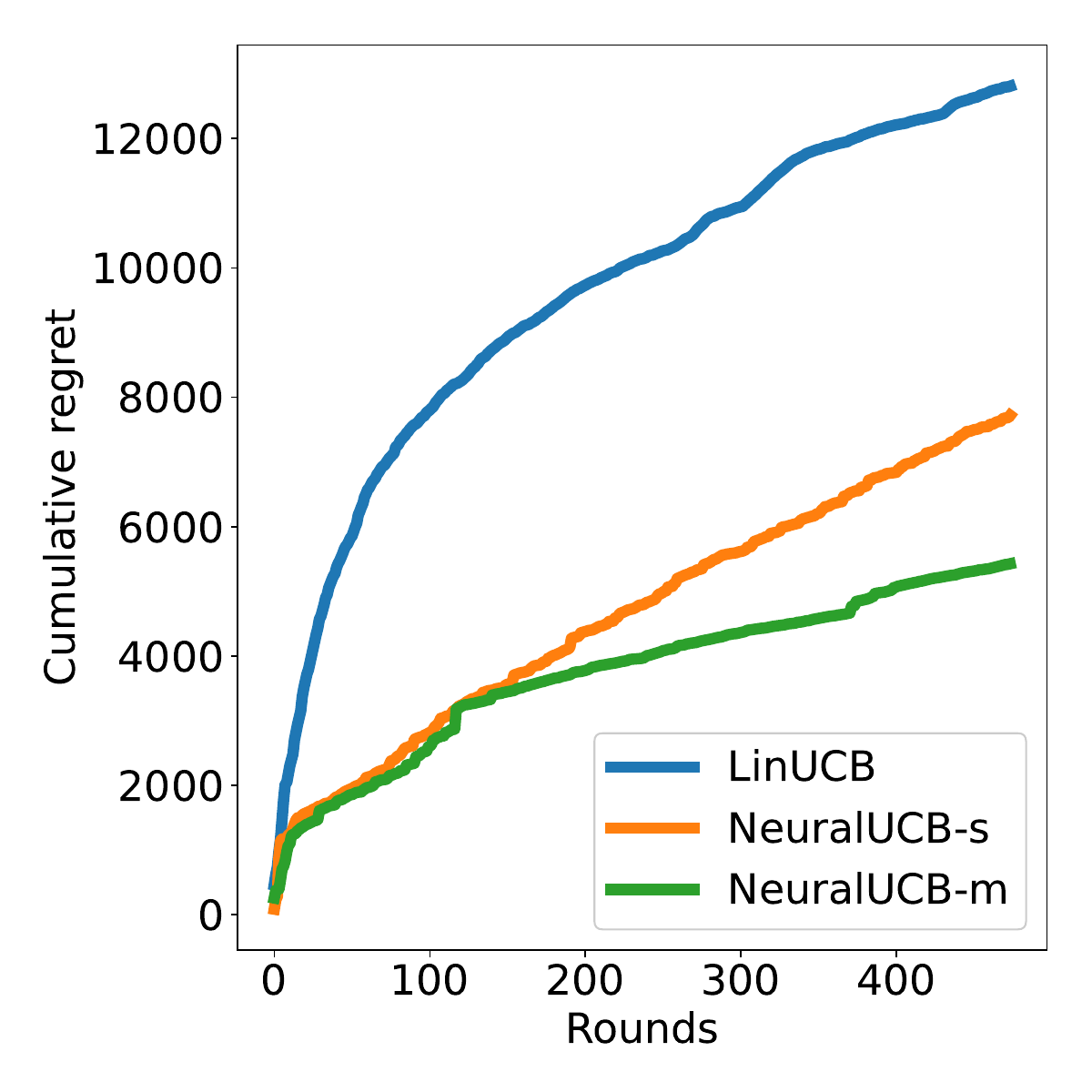}
    \vspace{-0.25cm}
	\caption{Comparison of UCB based client selection algorithms}
	\label{fig:regretvsrounds}
\end{figure}

\begin{table*}[]
\caption{Client selection evaluation}
\label{tab:client_selection_evaluation}
\resizebox{\textwidth}{!}{\begin{tabular}{|c|c|c|cccccccc|c|c|c|}
\hline
Approach                                                                            & Experiments                   & Clients  & \multicolumn{1}{c|}{$AC$}  & \multicolumn{1}{c|}{$BS$} & \multicolumn{1}{c|}{$e_{min}$} & \multicolumn{1}{c|}{$e_{max}$} & \multicolumn{1}{c|}{$\hat{b\_t}_{t,i}$} & \multicolumn{1}{c|}{$b_{max_{t,i}}$} & \multicolumn{1}{c|}{$e_{max_{t,i}}$} & \begin{tabular}[c]{@{}c@{}}$m_t$\\ (minutes)\end{tabular} & $e_{t,i}$ & \begin{tabular}[c]{@{}c@{}}Actual time\\ per  batch (seconds)\end{tabular} & \begin{tabular}[c]{@{}c@{}}Waiting time\\ (minutes)\end{tabular} \\ \hline
\multirow{4}{*}{\begin{tabular}[c]{@{}c@{}}Our \\ Client Selection\end{tabular}}    & \multirow{2}{*}{Scenario 1} & Client 1 & \multicolumn{1}{c|}{100} & \multicolumn{1}{c|}{1}  & \multicolumn{1}{c|}{1}    & \multicolumn{1}{c|}{7}    & \multicolumn{1}{c|}{431.93}  & \multicolumn{1}{c|}{46}        & \multicolumn{1}{c|}{7}         & \multirow{2}{*}{146.57}                                & 4      & 430                                                                        & \multirow{2}{*}{7.42}                                            \\ \cline{3-10} \cline{12-13}
&                               & Client 2 & \multicolumn{1}{c|}{100} & \multicolumn{1}{c|}{1}  & \multicolumn{1}{c|}{1}    & \multicolumn{1}{c|}{7}    & \multicolumn{1}{c|}{251.25}  & \multicolumn{1}{c|}{46}        & \multicolumn{1}{c|}{7}         &                                                        & 7      & 233                                                                        &                                                                  \\ \cline{2-14} 
& \multirow{2}{*}{Scenario 2} & Client 1 & \multicolumn{1}{c|}{60}  & \multicolumn{1}{c|}{0}  & \multicolumn{1}{c|}{1}    & \multicolumn{1}{c|}{7}    & \multicolumn{1}{c|}{251.25}  & \multicolumn{1}{c|}{18}        & \multicolumn{1}{c|}{3}         & \multirow{2}{*}{51.86}                                 & 3      & 233                                                                        & \multirow{2}{*}{14.25}                                           \\ \cline{3-10} \cline{12-13}
&                               & Client 2 & \multicolumn{1}{c|}{100} & \multicolumn{1}{c|}{0}  & \multicolumn{1}{c|}{1}    & \multicolumn{1}{c|}{7}    & \multicolumn{1}{c|}{130.36}  & \multicolumn{1}{c|}{50}        & \multicolumn{1}{c|}{7}         &                                                        & 4      & 132                                                                        &                                                                  \\ \hline
\multirow{4}{*}{\begin{tabular}[c]{@{}c@{}}Random \\ Client Selection\end{tabular}} & \multirow{2}{*}{Scenario 1} & Client 1 & \multicolumn{8}{c|}{\multirow{4}{*}{X}}                                                                                                                                                                                                                              & 7      & 430                                                                        & \multirow{2}{*}{114.92}                                          \\ \cline{3-3} \cline{12-13}
 &                               & Client 2 & \multicolumn{8}{c|}{}                                                                                                                                                                                                                                                & 7      & 233                                                                        &                                                                  \\ \cline{2-3} \cline{12-14} 
& \multirow{2}{*}{Scenario 2} & Client 1 & \multicolumn{8}{c|}{}                                                                                                                                                                                                                                                & 7      & 233                                                                        & \multirow{2}{*}{$\infty$}                                        \\ \cline{3-3} \cline{12-13}
&                               & Client 2 & \multicolumn{8}{c|}{}                                                                                                                                                                                                                                                & 7      & 132                                                                        &                                                                  \\ \hline
\end{tabular}}
\vspace{-0.3cm}

\end{table*}

\subsection{Effect of resources on the training time}
\label{sec:resources_effect}
Figure \ref{fig:ram_effect} depicts the results of an experiment to show the effect of varying RAM on training time with the help of two scenarios: one with background apps  running alongside our FL android application (less $AR$) and other with no background apps (high $AR$). We observe that with decrease in available RAM, there is a significant increase in training time per batch, across all the mobile phones. This is especially noticeable in Figure \ref{fig:1_ram} and Figure \ref{fig:4_ram}, where we see a jump of 49 and 33 seconds in training time respectively. Figure \ref{fig:battery_effect} presents the results obtained during the experiment conducted to check the effect of the battery percentage on the phone's training time.
We can infer from the figures that training time shoots up abruptly when in lower battery bands ($\gamma= 20\%$), whereas it is almost constant in the upper battery bands for all the phones except Xiaomi 11 Pro. This is particularly evident in the OnePlus 5T phone where training time increased 2.4 times the regular time in lower battery bands.

\subsection{Neural reward generator}
\label{sec:results_neural_reward_generator}
For our contextual bandits experiment, we chose $N=4$ clients and the number of rounds $T= 475$. We ran multiple iterations of on-device training on the $N$ mobile devices to generate the context vectors containing the resource information and noted down the time taken per batch and the battery drop. We use the experimental setup described in \cite{linucb} for LinUCB. The neural networks used in the NeuralUCB-s and NeuralUCB-m share the same architecture, consisting of a simple fully connected feedforward network with two hidden layers of 32 and 16 units, respectively with ReLu activations.  The input to the network is a d dimensional context vector, and the outputs are the time and battery drop. For NeuralUCB-s and LinUCB, we use a single reward generating function for all the clients. As a result the context vector $c=\left[TR,AR,AC,BS,CI,PI\right]$ contains all six features as discussed in Section \ref{resources}. In contrast, we do not need to provide the total RAM (TR) and phone-specific information (PI) as features in the NeuralUCB-m approach with personalised models for each client device. Hence, $d=4$ is used in this case. We do a grid search over $\{ 0.01,0.1,1.0,10\}$ for the exploration term multiplier $\alpha_t=\alpha$ for all the experiments, and tuned the parameter in such a way that the fairness achieved is similar across all the algorithms. For LinUCB, we selected $\alpha=10.0$ and a value of $0.01$ for NeuralUCB based algorithms.

Figure \ref{fig:mse_plots} trace the model's mean square error loss over rounds. We can see that neural network-based algorithms outperform LinUCB. Figure \ref{fig:linucbmse} demonstrates that linear models are unable to accurately estimate the output and learn feature representations from the data. Comparing the results of NeuralUCB-s and NeuralUCB-m from Figures \ref{fig:neuralucb-s} and \ref{fig:neuralucb-m} respectively, we can observe that both loss curves looks similar with NeuralUCB-m performing slightly better over a long run. Figure \ref{fig:regretvsrounds} plots the overall regret of all the algorithms against rounds. We present the findings based on an average of five repeated experiments using various random dataset shuffles. Further, we can also infer from Figure \ref{fig:regretvsrounds} that NeuralUCB-m with disjoint personalised models for each client appears to outperform all other algorithms.

\begin{figure}
\begin{minipage}[t]{0.485\columnwidth}
  \includegraphics[width =\linewidth]{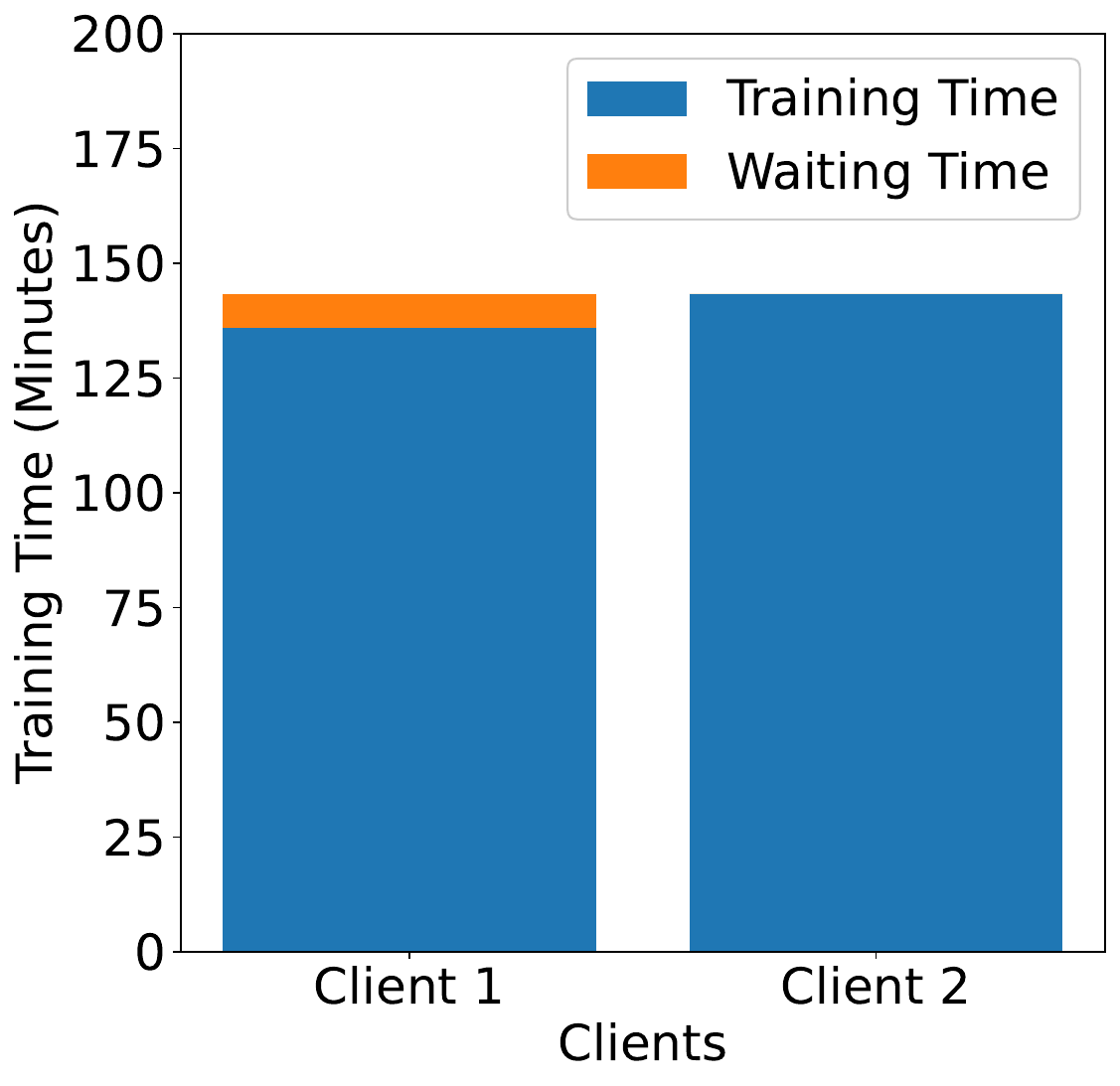}
    \caption{Scenario 1: Slow vs Fast client.}
  \label{fig:our_approach_waiting_time1}
\end{minipage}
\begin{minipage}[t]{0.485\columnwidth}
  \includegraphics[width = \linewidth]{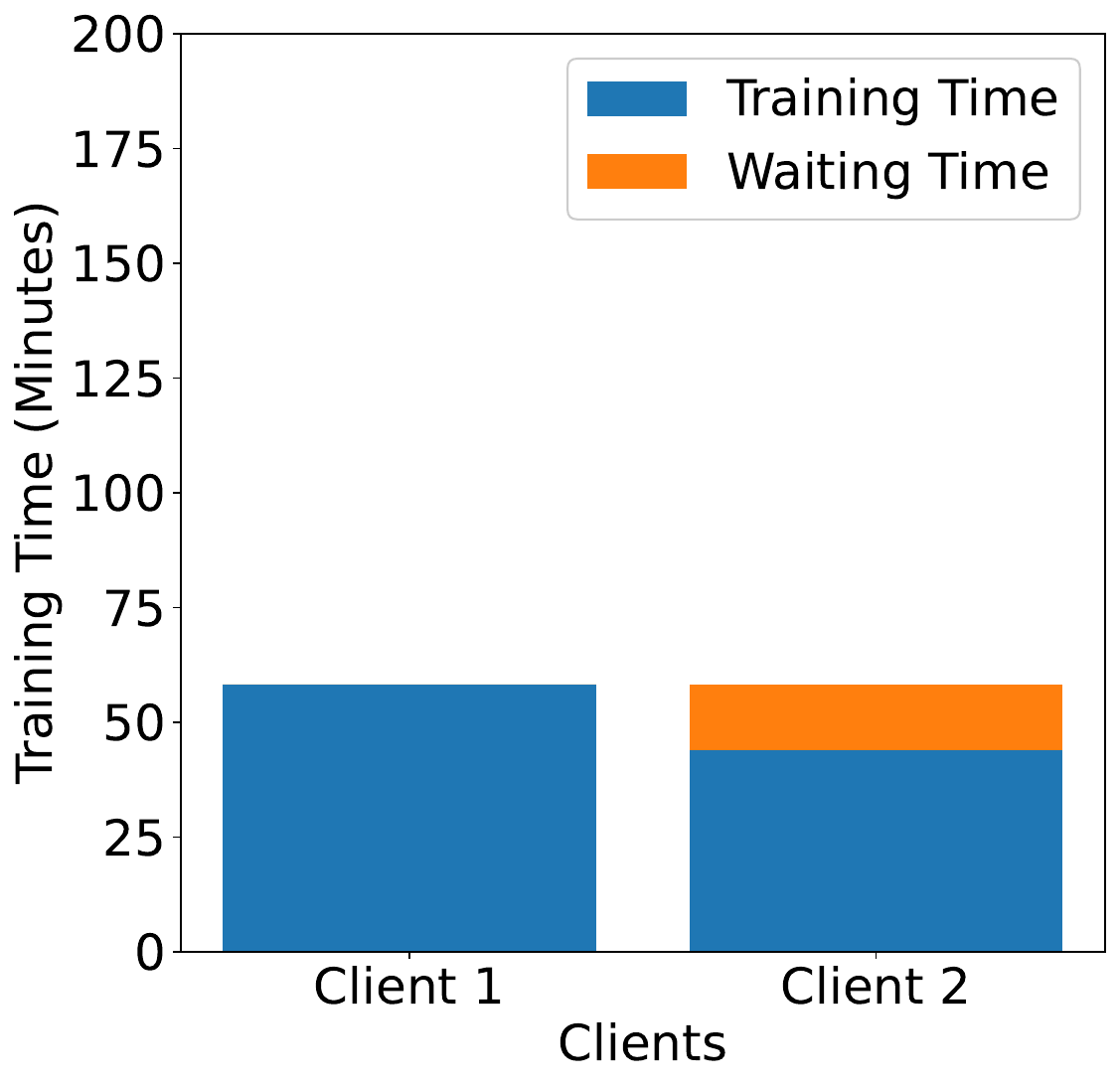}
  \caption{Scenario 2: One client with insufficient battery life.}
  \label{fig:our_approach_waiting_time2}
\end{minipage}\hfill % maximize horizontal separation
\end{figure}

\begin{figure}
\begin{minipage}[t]{0.495\columnwidth}
  \includegraphics[width=\linewidth]{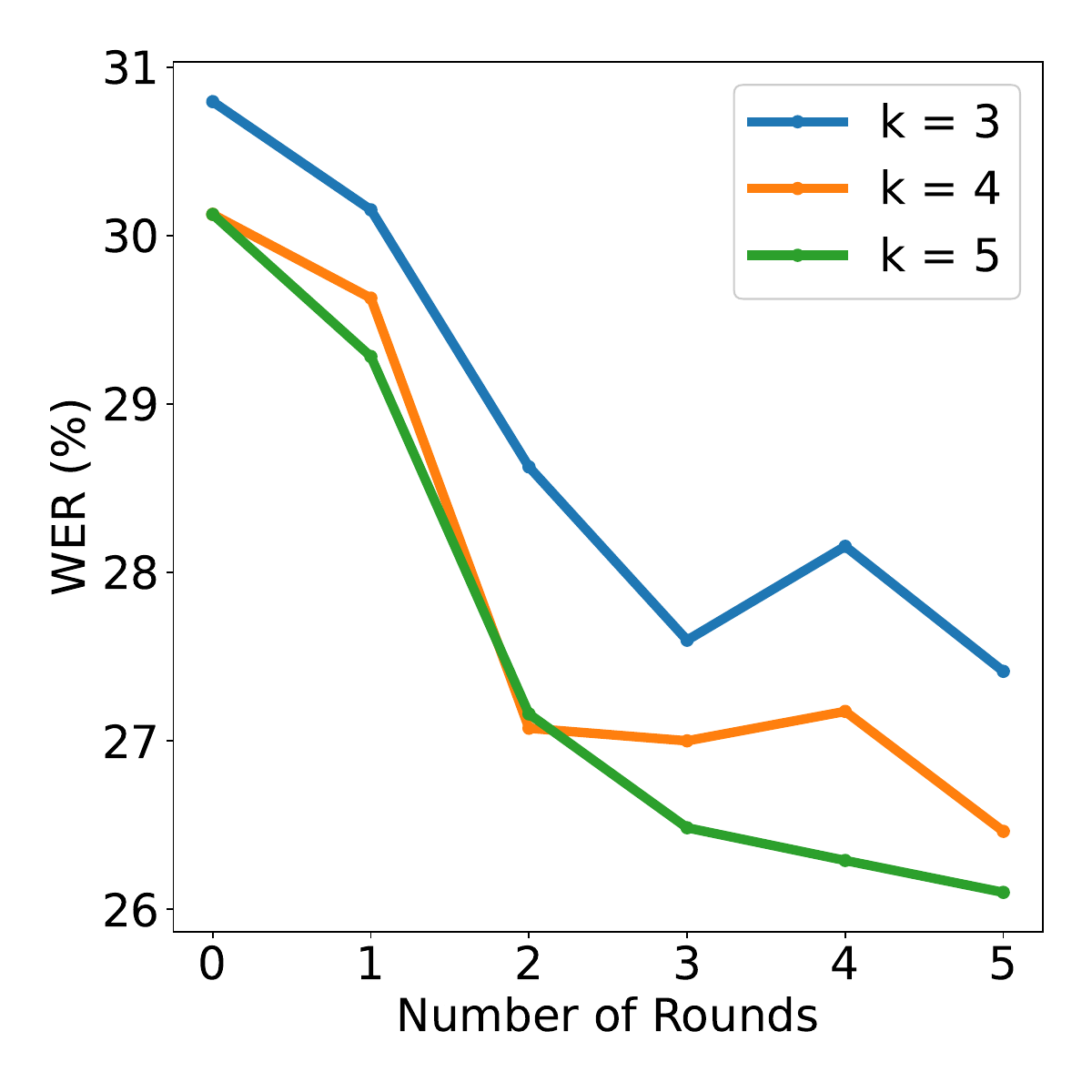}
  \caption{Trend of WER for various values of k}
  \label{fig:multiple_clients}
\end{minipage}\hfill % maximize horizontal separation
\begin{minipage}[t]{0.495\columnwidth}
  \includegraphics[width=\linewidth]{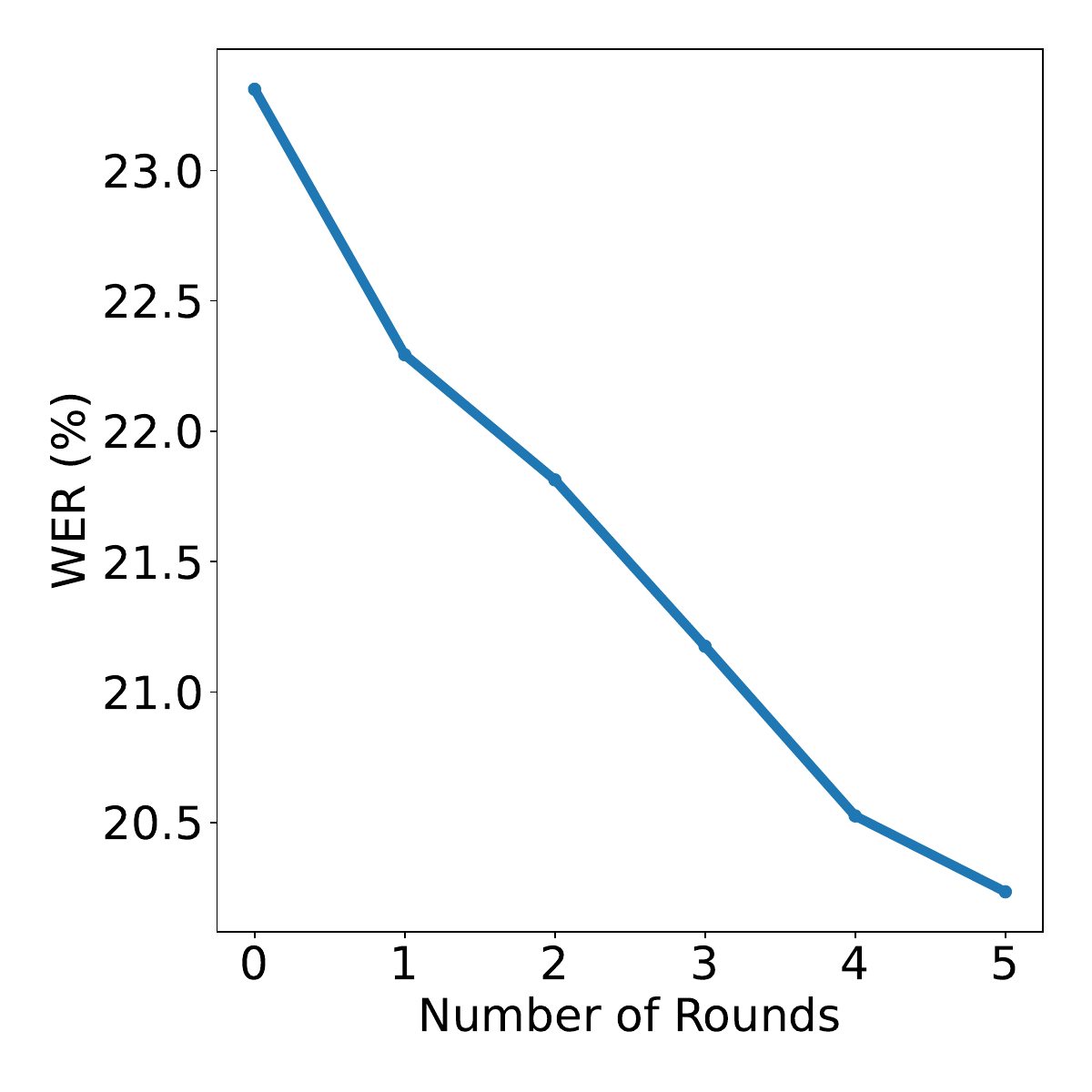}

  \caption{Performance of Ed-Fed on mobile phones}
\label{fig:mobile_results}

\end{minipage}
\vspace{-0.5cm}

\end{figure}

\subsection{Resource-aware time-optimised client selection}
Figure \ref{fig:our_approach_waiting_time1} and Figure \ref{fig:our_approach_waiting_time2} are the results obtained when we redo the experiments of Scenario 1 and Scenario 2 of Section \ref{sec:need_for_waiting_time} with our client selection algorithm using our neural reward generator at $t = 476$.
Table \ref{tab:client_selection_evaluation} contains the detailed information of these two experiments. 
In Scenario 1, our algorithm identified that client 1 has weaker computing capabilities based on the $\hat{b\_t}_{t,i}$ values computed. Since, client 1 has weaker computing capabilities, our algorithm assigns a smaller $e_{t,1}$ of 4 epochs in contrast to the 7 epochs given by random client selection approach. Thereby, reducing the overall waiting time to 7.42 minutes when compared to random client selection's 114.92 minutes.
In Scenario 2, it is clear from $e_{max_{t,i}}$ values of the Table \ref{tab:client_selection_evaluation} that client 1 has weak battery resources. Our client selection algorithm deduced that client 1 cannot run 7 epochs and prevented the whole FL round from stopping by assigning a smaller $e_{t,1}$ of 3 epochs unlike random client selection. Further, our algorithm adjusted $e_{t,2}$ with respect to $e_{t,1}$ by giving it a value of 4 and reduced the waiting time even more. Whereas, the random client selection without any knowledge of resource information asks the client 1 to run $e_{max}$ epochs. This leads to power shutdown of client 1, thereby making client 2 wait for infinite amount of time. 

\subsection{Ed-Fed framework evaluation}

Figure \ref{fig:multiple_clients} displays the results obtained by choosing various values of k for each FL experiment. The figure shows three line plots, each corresponding to different values of k. Each line plot shows how the global model performs on the global test set after each round of FL. The global test set consists of 15 unseen speech samples each from 10 speakers with 4 different accents. We can deduce from the figure that as k increases, the WER of the global model decreases.

Figure \ref{fig:mobile_results} depicts the findings obtained on deployment of our Ed-Fed framework on multiple phones. The experiment is carried for 5 rounds on 4 mobile devices. In each round, 2 clients are selected. The round 0 in the figure refers to the initial global weights. All the checkpoints that are obtained at the end of each FL round are put to the test on a global test set. As could be predicted, the WER declines as the number of FL rounds grow.

\section{CONCLUSIONS AND FUTURE DIRECTIONS}
\label{sec:conclusion}
In this work, we present Ed-Fed, a first-of-its-kind end-to-end federated learning framework that will serve as a foundation for future research in practical FL systems. We also propose a client selection algorithm that takes into account factors such as computation, storage, power, and device-specific capabilities to handle stragglers, optimise waiting time, and adapt training time based on resource information. 
The framework has been thoroughly tested in simulations and on actual edge devices,  and in the future, we plan to incorporate communication latency parameters to measure waiting time.

\end{document}